\documentclass[iop]{emulateapj}
\usepackage{color} 
\usepackage{epsfig}
\usepackage{graphicx}

\newcommand{\kms} {{km}\hskip1pt{s$^{\scalebox{.6}{-1}}$}}
\newcommand{\kmps} {{km}\hskip1pt{s$^{-1}$}}
\newcommand\M{\rule{0pt}{2.3ex}}       

\shorttitle{Images of Gravitational and Magnetic Phenomena in Interacting Binary Stars}
\shortauthors{Richards, Cocking, Fisher, \& Conover}

\begin{document}

\title{Images of Gravitational and Magnetic Phenomena Derived from 2D Back-Projection Doppler Tomography of Interacting Binary Stars }

\author{Mercedes T. Richards\altaffilmark{1,2}, Alexander S. Cocking\altaffilmark{1}, John G. Fisher\altaffilmark{1} \& Marshall J. Conover\altaffilmark{1}}
\affil{
{$^1$}Department of Astronomy \& Astrophysics, Pennsylvania State University, University Park, PA 16802, U.S.A. \\
mrichards@astro.psu.edu, asc5097@psu.edu \\
{$^2$}Institut f\"ur Angewandte Mathematik, Ruprecht-Karls-Universit\"at Heidelberg, 69120 Heidelberg, Germany\\
} 

\begin{abstract} 

We have used 2D back-projection Doppler tomography as a tool to examine the influence of gravitational and magnetic phenomena in interacting binaries which undergo mass transfer from a magnetically-active star onto a non-magnetic main sequence star. This multi-tiered study of over 1300 time-resolved spectra of 13 Algol binaries involved calculations of the predicted dynamical behavior of the gravitational flow and the dynamics at the impact site, analysis of the velocity images constructed from tomography, and the influence on the tomograms of orbital inclination, systemic velocity, orbital coverage, and shadowing.  The H$\alpha$ tomograms revealed eight sources: chromospheric emission, a gas stream along the gravitational trajectory, a star-stream impact region, a bulge of absorption or emission around the mass-gaining star, a Keplerian accretion disk, an absorption zone associated with hotter gas, a disk-stream impact region, and a hot spot where the stream strikes the edge of a disk. We described several methods used to extract the physical properties of the emission sources directly from the velocity images, including  S-wave analysis, the creation of simulated velocity tomograms from hydrodynamic simulations, and  the use of synthetic spectra with tomography to sequentially extract the separate sources of emission from the velocity image.  In summary, the tomography images have revealed results that cannot be explained solely by gravitational effects: chromospheric emission moving with the mass-losing star, a gas stream deflected from the gravitational trajectory, and alternating behavior between stream state and disk state.  Our results demonstrate that magnetic effects cannot be ignored in these interacting binaries.

\end{abstract}


\section{Introduction}            

The effects of gravitational forces and magnetic fields have been recognized as playing an important role in the mass accretion process in a variety of astrophysical systems, including  protostars, binaries containing compact objects, and massive galaxies with supermassive black holes.  However, the consequences for accretion onto normal stars in close binary star systems are not as well understood.  In this paper, we focus on binaries containing normal stars, specifically the Algol-type binaries with prototype $\beta$ Per (Algol), which are progenitors of the cataclysmic variables \citep{iben+tutukov85,wheeler07}.   In these systems, the cool star displays enhanced magnetic activity associated with rapid rotation induced by tidal synchronization of the binary system, and the resulting dynamic phenomena are many times more powerful than those found on the Sun. Since the cool star is the source of the mass transfer stream, then this enhanced activity will influence the mass transfer process in a direct way.

In the Algols, no magnetic activity is linked to the hot B-A-type mass-gaining star \citep{richards+albright93,petersonetal10}.  However, both gyrosynchrotron radio emission and thermal X-ray emission are expected from the cool F-K-type mass-losing companion, mostly due to coronal activity \citep{richards+albright93}.   Additional evidence of magnetic activity can be derived from the spectra (e.g., CaII H \& K or H$\alpha$ emission), however such evidence is typically sparse because these cool stars contribute only $\sim$ 5-15{\%} to the total luminosity of the binary, depending on wavelength, with the highest contribution in the infrared.  Moreover, any magnetic components in the H$\alpha$ line are masked by contributions from circumstellar gas flowing between and around the stars \citep{richards+albright93}.   These magnetic phenomena observed in Algol binaries are similar to those detected in the double-cool star detached RS CVn binaries \citep{guinan+gimenez94}.   

The presence of an active magnetic field on the mass-losing star can lead to magnetic threading of the gas flow from the L1 point towards the mass-gaining star when the ram pressure of the accretion flow exceeds the magnetic stress at the L1 point \citep{retteretal05}.  As a consequence, the magnetic field will be dragged along with the fluid flow and eventually surround the mass gainer.  This process is associated with the superhump phenomenon, which typically arises from the prograde apsidal precession of an eccentric accretion disk and is usually identified in cataclysmic variables \citep{vogt82, patterson98,patterson99}.  Superhumps were first detected in Algol binaries by \citet{retteretal05} using 2.3 GHz and 8.3 GHz observations of $\beta$ Per derived from the 5.6-year continuous radio flare survey of \citet{richards+waltmanetal03}.  This prediction of the superhump  phenomenon in $\beta$ Per was confirmed by \citet{petersonetal10} using 15 GHz VLBI observations.  Astrometric methods have also been used to confirm the association between the source of the radio emission and the physical location of the K-type mass-losing star, with a precision of $\pm$0.5 mas \citep{zavalaetal10,petersonetal11}.

\begin{figure*}[t]
\begin{center}
\epsscale{0.95}
\plotone{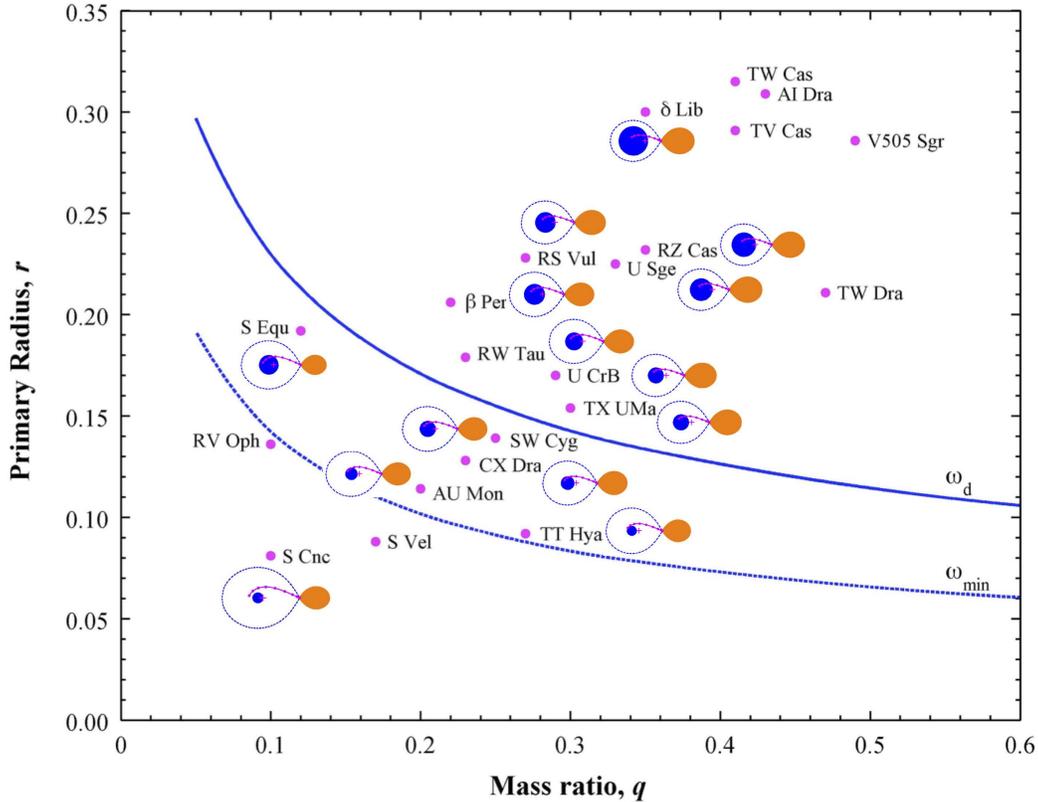}
\end{center}
\vspace{-0.5cm}
\caption{The {\it r--q} diagram, where {\it r} is the radius of the primary star in units of the separation of the binary and {\it
q} is the mass ratio. The two curves plotted $\omega_{d}$ and $\omega_{min}$ represent the smallest radius of a stable accretion disk and the distance of closest approach of the gas stream, respectively.
}
\label{fig1}
\end{figure*}

Indirect images of interacting binaries created with the aid of Doppler tomography have led to a better understanding of the role played by gravitational forces and magnetic fields.   This image reconstruction technique was introduced by \citet{marsh+horne88} and then applied to cataclysmic variables and X-ray binaries \citep{kaitchucketal94,morales-rueda04,schwopeetal04,steeghs04,vrtileketal04} and also to direct-impact and long-period Algol binaries \citep{richards+albright+bowles95,richards+albright96,richards04}.  The images illustrate the movement of gas from the inner Lagrangian point towards and around normal or compact mass-gaining stars, and show that a well-defined accretion disk is formed around the mass-gaining star in the compact binaries.   However, in the {\it direct-impact} systems the mass gainer is a normal main sequence star, and the stars are very close to each other relative to the orbital separation.  The first 2D tomography images of these latter systems have shown that the path of the flow is usually governed by gravitational and Coriolis forces, and there is additional evidence that the active cool mass-losing star can have substantial influence on the mass transfer process \citep{richards+albright96,richards+jones+swain96}. 

In this paper, we assess the progress that has been made in 2D tomography of the direct-impact and long-period Algols to identify patterns and similarities between systems with orbital periods from 1.2 days to 11.1 days; some preliminary results have appeared in conference proceedings (e.g., \citealt{richards04}).  In \S2,  we describe the predicted dynamical behavior of the gravitational flow along the gas stream and beyond; in \S3, we describe the technique of Doppler tomography, and \S4 provides the application to the direct-impact systems.   We also examine the influence of various system properties on the 2D images in \S5, specifically the effects of orbital inclination, systemic velocity, orbital phase coverage, inadequate velocity resolution, and the effect of the gas opacity known as shadowing.   In \S6, we describe the techniques that can be used to extract physical properties of the gas flows directly from the images via  four methods: an S-wave analysis; using hydrodynamic simulations to create simulated tomograms; using synthetic spectra to model the contributions of the accretion disk, gas stream, and other sources of emission; and using synthetic spectra with tomography to confirm visually that the separate accretion structures have been modeled accurately.   The Results and Conclusions are given in \S7 and \S8, respectively.  

\section{Understanding the Direct-impact Systems}

The semidetached interacting Algol binaries contain a cool F-K III-IV secondary star which has expanded to fill its Roche lobe and is transferring material through a gas stream onto a hot B-A V primary star (the mass gainer); in this context, the primary is the brighter and more massive star.  A radius vs. mass ratio or $r-q$ diagram, as illustrated in Figure \ref{fig1}, can be used to predict the systems that should undergo direct impact of the gas stream onto the stellar photosphere.  This figure also shows the Roche geometries for the direct-impact systems and some wider binaries based on the properties listed in Table \ref{tab1}, derived from \citet{richards+albright99}. Here, $P$ is the orbital period of the binary; $i$ is the orbital inclination; $q=M_2/M_1$ is the mass ratio of the binary; $r = R_1/a$ is the radius of the mass gainer, $R_1$, relative to the binary separation $a$; $K_1$  and $K_2 (= K_1/q)$ are the velocity semi-amplitudes of the primary and secondary, respectively; and $V_o$ is the systemic velocity of the binary.  Until these systems are resolved by an instrument with absolute phase angle calibration or direct imaging, the values of inclination will have an $i$ vs. $(180-i)$ ambiguity, corresponding to prograde vs. retrograde orbits.  For the close binary in $\beta$ Per, \citet{zavalaetal10} found that the orbit of the binary is retrograde, so the inclination should be $(180 - 81.4) = 98.6^\circ$.   However, we retain the usual convention here with angles up to $90 ^\circ$ corresponding to prograde orbits. 

In Figure \ref{fig1}, the direct-impact binaries will be found above the solid curve labeled $\omega_{d}$, corresponding to the radius of a stable accretion disk (e.g., $\delta$ Lib, RZ Cas, U Sge) since the mass gainer is larger in size than the predicted disk radius. However, the wider binaries that always form a stable accretion disk (e.g., S Cnc) will be found below the dashed curve labeled $\omega_{min}$, corresponding to the distance of closest approach of the gas stream, hence no impact is predicted since the size of the star is smaller than $\omega_{min}$; these curves are based on the semi-analytical ballistic calculations of \citet{lubow+shu75}.   It is noticeable that systems between the two curves may have a grazing impact between the gas stream and the mass gainer (e.g., S Equ, SW Cyg, CX Dra) while no impact is expected in other cases (e.g., TT Hya).  In this scenario, the direct-impact systems typically have $P < 5-6$ days \citep{struve49} while the grazing-impact systems have $6 < P < 12$ days. 

\begin{table}[t]
\caption{Orbital Parameters}
\centering
\begin{tabular}{lcccrcr}
\hline\hline \M
System & $P$ &   $i$         & $q$ & $R_1/a$ & $K_1$            & $V_o$~~~ \\
{}           & (days) & ($^\circ$) &        &                &  (\kms) & (\kms)            \\[0.5ex]\hline \M
RZ Cas        & 1.195 & 83.0 & 0.33 & 0.232   & 70.9 & $-$45.5 \\  
$\delta$ Lib & 2.327 & 78.6 & 0.35 & 0.300   & 76.6 & $+$~4.0 \\
RW Tau       & 2.769 & 90.0 & 0.23 & 0.179    & 53.0 & $-$20.2 \\
$\beta$ Per & 2.867 & 81.4 & 0.22 & 0.206	& 44.0 & $+$~3.8 \\
TX UMa      & 3.063 & 81.0 & 0.30 & 0.154	& 54.8 & $-$13.0 \\ 
U Sge         & 3.381 & 89.0 & 0.33 & 0.225	& 70.0 & $-$10.0 \\ 
S Equ         & 3.436 & 87.4 & 0.12 & 0.192	& 23.0 & $-$48.0 \\
U CrB         & 3.452 & 79.1 & 0.29 & 0.170	& 59.7 & $-$~6.7 \\ 
RS Vul        & 4.478 & 78.7 & 0.31 & 0.267	& 54.0 & $-$20.1 \\  
SW Cyg      & 4.573 & 87.1 & 0.21 & 0.155     & 32.3  & $-$~7.5 \\  
CX Dra       & 6.696 & 53.0 & 0.23 & 0.128	& 34.1 & $+$~3.54\\
TT Hya       & 6.953 & 84.4 & 0.27 & 0.089	& 23.9  & $+$~0.0 \\  
AU Mon      & 11.113~ & 82.0 & 0.20 & 0.114	& 32.2 & $+$11.8 \\  
V711 Tau     & 2.838 & 33.0 & 1.25 &  0.113	& 61.7 & $-$14.01\\
\hline
\end{tabular}
\label{tab1}
\end{table}

The dynamics of the direct impact can be understood in terms of the parameters given in Table \ref{tab2}, which  lists
the thermal or Kelvin-Helmholtz timescale,  $t_{kh} = G M^2/R L$, where $L = 4\pi R^2 \sigma T^4$; 
the dynamical or free fall timescale, $t_{dyn} \simeq (R^3/GM)^{0.5}$; 
the circularization timescale or the time for the gas to circle the mass gainer at a distance $d$ from the center of the star, $t_{cir}=2\pi d/v_{cir}$, where $v_{cir} =  (GM/d)^{0.5}$ and the values in the table correspond to $d=R$;
the synchronous velocity of each binary, $v_{syn} = 2\pi R/P$; 
the escape velocity for the mass gainer, $v_{esc} = (2GM/R)^{0.5}$; and 
the velocity of the gas stream at the impact site or shock region, $v_{imp}$, derived from the calculation of the gas stream trajectory.
For these calculations, the mass, radius, luminosity, and temperature all refer to the mass-gaining star, and $P$ is the orbital period.

The values of $t_{kh}$ in Table \ref{tab2} suggest that the mass gainer in a typical direct-impact system would radiate its reserves of thermal energy within $\sim 0.5 - 1.8 \times 10^6$ yr.  However, the star would take only $\sim$ 0.8 - 3.9 hr to respond to the inflow of gas from the companion star ($t_{dyn}$ in Table \ref{tab2}).  Moreover, the gas beyond the impact site would circle the star in only  $\sim 4 - 17$ hr at the stellar surface, $d=R$ (see $t_{cir}$ in Table \ref{tab2}) or $\sim 7 - 32$ hr at a distance $d = 1.5R$ from the center of the star.  Moreover, the impact of the gas should spin up the star since the stream impact velocities of $430 - 740$ {\kmps} are much higher than the $\sim 15 - 90$ {\kmps} synchronous velocities of the mass gainer.  The stream impact velocity is lower than the predicted escape velocity ($\sim 550 - 830$ {\kmps}) except in the cases of RW Tau, RS Vul, and AU Mon.    

\begin{table}[t]
\setlength{\tabcolsep}{4pt}
\caption{Dynamical Parameters}
\centering
\begin{tabular}{lcccccc}
\hline\hline \M
System &  $t_{kh}$  &   $t_{dyn}$  &   $t_{cir}$  & $v_{syn}$   &  $v_{esc}$ & $v_{imp}$    \\
{}           & ($10^6$yr)  & (hr)           &  (hr)          &  (\kms)        &  (\kms)            & (\kms)        \\
[0.5ex]\hline \M
RZ Cas	&	1.620	&	0.84	&	~3.75	&	62.3	&	673.8	&	514 $\pm$ 9~	\\
$\delta$ Lib	&	0.530	&	2.42	&	10.73	&	89.6	&	659.6	&	434 $\pm$ 3~	\\
RW Tau	&	1.070	&	1.27	&	~5.63	&	40.2	&	671.3	&	739 $\pm$ 11	\\
$\beta$ Per	&	0.943	&	1.61	&	~7.14	&	51.2	&	697.5	&	557 $\pm$ 7~ \\
TX UMa	&	1.790	&	1.15	&	~5.11	&	38.0	&	772.6	&	644 $\pm$ 11	\\
U Sge	&	0.736	&	2.26	&	10.03	&	62.9	&	719.4	&	558 $\pm$ 13	\\
S Equ	&	0.735	&	1.67	&	~7.40	&	41.2	&	649.8	&	566 $\pm$ 11	\\
U CrB	&	1.430	&	1.49	&	~6.60	&	44.0	&	781.1	&	719 $\pm$ 22	\\
RS Vul	&	0.195	&	3.85	&	17.11	&	62.2	&	552.3	&	581 $\pm$ 8~	\\
SW Cyg	&	0.597	&	1.66	&	~7.38	&	28.8	&	605.5	&	534 $\pm$ 8~		\\
CX Dra	&	1.400	&	1.85	&	~8.24	&	30.2	&	834.2	&	576 $\pm$ 17	\\
TT Hya	&	1.340	&	1.13	&	~5.03	&	14.8	&	697.2	&	-	\\
AU Mon	&	0.621	&	2.52	&	11.20	&	21.0	&	705.3	&	691 $\pm$ 65	\\
\hline
\end{tabular}
\label{tab2}
\end{table}

\begin{figure*}[t]
\begin{center}
\epsscale{1}
\plotone{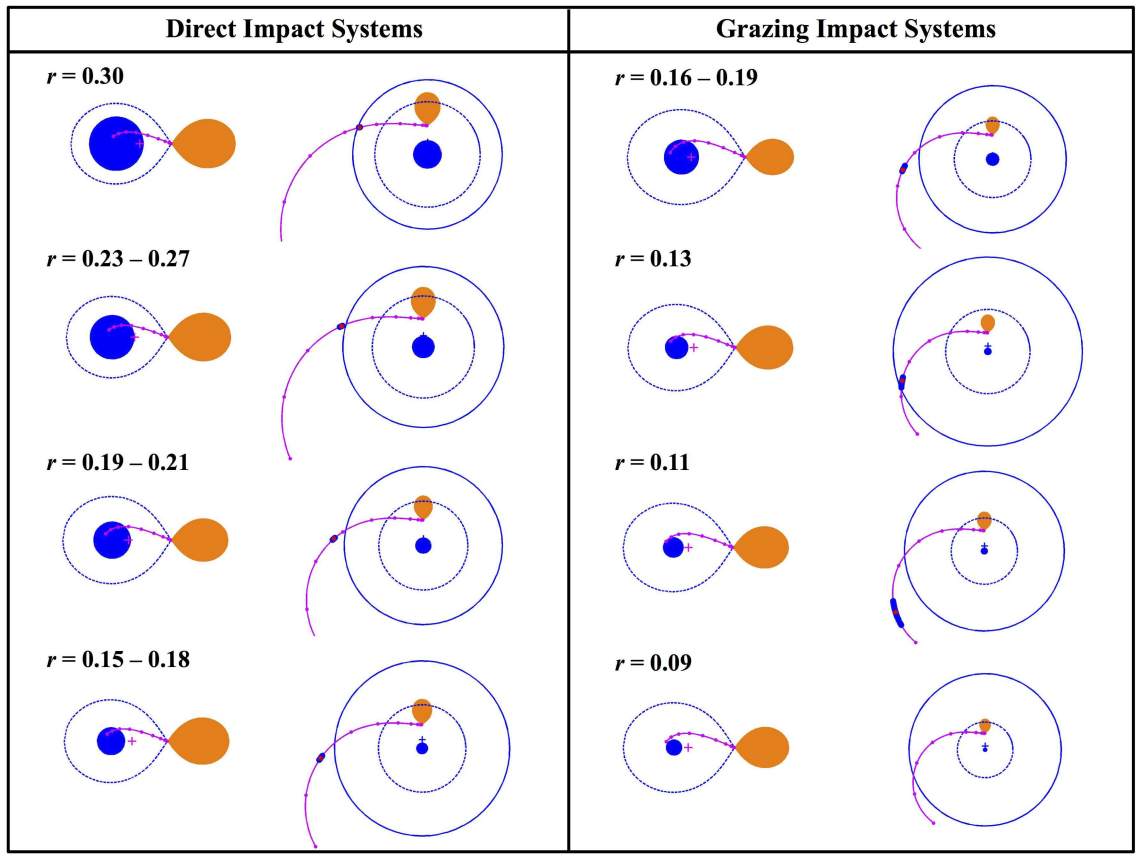}
\end{center}
\vspace{-0.4cm}
\caption{Predicted location of the impact site or shock region for direct-impact and grazing-impact systems, for decreasing values of $r = R_1/a$.  Each subsection shows  the Cartesian frame (left image) and velocity frame (right image).
}
\label{fig2}
\end{figure*}

To illustrate the expected behavior of the gas flow in the velocity domain, we compared plots of the Cartesian representation of the binary with the corresponding velocity frame (see Figure \ref{fig2}).   Since the direct-impact systems have eccentricities that are very close to zero, then the binaries can be assumed to be tidally locked. The assumption of synchronous rotation ($P_{orb}=P_{rot}$) provides a relationship between the length and velocity scales that permits the locations of the stars to be plotted in the velocity frame.
Figure \ref{fig2} also displays two large circles around the binary corresponding to the locus of a Keplerian accretion disk in the velocity frame if that disk fills the space between the photosphere and the Roche surface of the mass gainer; the largest solid circle and the smaller dashed circle mark the inner and outer edge of the accretion disk, respectively; and the plus sign marks the center of mass of the binary.  

In both the Cartesian and velocity frames, the solid trajectory is the gravitational free-fall path of the gas stream from the L1 point under the influence of the Coriolis force, in the frame of reference of the binary. The trajectory has been divided into ten parts marked by small solid circles to permit a visual comparison of the simultaneous position and velocity of a specific location along the gas stream trajectories.
In addition, the large solid red circle and the extended blue region along the trajectory represent the median and range, respectively, of the predicted location of the stream-photosphere impact site.  The values of the stream velocity at the impact site are listed under $v_{imp}$ in the last column of Table \ref{tab2}.

The impact site was examined in greater detail by \citet{richards92} by calculating the ram pressure in the gas stream and the depth of the shock region relative to the stellar photosphere.  The depth at which the shock forms is sensitive to the assumed valid of the mass transfer rate. Over a range of mass transfer rates, the stream reached a depth of $2 \times 10^{-5} R_1$ below the photosphere to $4 \times 10^{-6} R_1$ above the photoshere in the case of $\beta$ Per, suggesting that most of the flow is reflected away from the shock region in this binary.  The tangential component of the impact velocities listed in Table \ref{tab2} can produce substantial differential rotation and redistribution of the gas flow around the mass gainer; and observational evidence of such high rotational velocity regions was found in ultraviolet spectra by \citet{cugier+molaro84}  and at H$\alpha$ by \citet{richards93}.  The impact also produces heating of the gas to temperatures of $\sim 10^{5-6}$ K, and evidence of the shock region was detected as a high temperature accretion region in ultraviolet spectra \citep{peters+polidan84}.  Similar behavior is expected for the other direct-impact systems.

So, purely in terms of gravitational and Coriolis forces, the gas flow from the L1 point will make direct impact with the surface of the mass gainer in most of the systems under study.  Since the flow velocities are much higher than the synchronous velocity of the star, the impact should create an equatorial bulge on the surface of the star, and in a few cases, the gas may be expelled from the system. The conditions at the impact site will be turbulent, and the tangential component of the flow will propel the gas around the star well beyond the impact site on timescales of hours to days, until it reaches the location of the incoming gas stream. 

\section{Impact of Doppler Tomography}

The image reconstruction technique of tomography can be implemented in two steps: (1) the acquisition of a set of views of the object at directions around the object over the 360$^{\circ}$ range of angles; these views are called slices or projections that can be represented mathematically by the Radon transform \citep{radon1917,shepp83}. (2) The recovery of an image of the original object may be obtained through a summation process called back projection, accomplished by taking each image projection and returning it along the path from which it was acquired.  The quality of the reconstruction depends on the number of views acquired and the angular distribution of these views.  In general, back-projection tomography uses ($n-1$)-dimensional projections to calculate an $n$-dimensional image of an object (e.g., construct a 3D image using 2D projections, like CAT scans in medicine).   Moreover, the technique is robust in that it can be applied to different types of measurements, for example: X-rays (medicine or archaeology), radar (satellite aperture radar - SAR), sonar (oceanography), seismic waves (geophysics), and spectra (astronomy).

Tomography can be applied only if there exist different angular views of a system. For example, since the surface of Venus is obscured by thick clouds and is opaque at optical wavelengths, the reconstruction of its surface was achieved at radio frequencies using radar data collected by the NASA Magellan satellite which orbited the planet for four years from 1990-1994.  Eclipsing binaries and rotating stars also provide us with changing views of the system and the reconstruction requires the use of spectra.  In these latter cases, the gas motions detected through Doppler shifts of spectral lines are used to provide an image of the gas flows in velocity coordinates, so the technique is termed {\it Doppler tomography}. 	

In two-dimensional back-projection Doppler tomography, the Radon transform of the function, $I(v_x,v_y)$, is a set of projections, $p(v,\phi)$ representing the {\it line profile}, with Doppler shift, $v_r$, at each orbital phase, $\phi$.
\begin{eqnarray*} 
p(v,\phi) &=& \hskip-2pt {\int^{\infty}_{-\infty}\hskip-1pt \int^{\infty}_{-\infty}} 
~I(v_x,v_y) \delta(v-v_r) ~dv_x~dv_y
\end{eqnarray*}
The reconstruction creates the 2D image $I(v_x,v_y)$ directly from the 1D Doppler shifts of the line profile by inverting the equation for the Radon transform \citep{marsh+horne88,kaitchucketal94}.
\begin{eqnarray*} 
I(v_x,v_y) &=& \hskip-2pt {\mathop{\int^{2\pi}_{0}\hskip-7pt {\int^{\infty}_{-\infty}\hskip-1pt \int^{\infty}_{-\infty}}}} 
~p(v,\phi) ~|\omega| ~e^{2\pi i\omega (v-v_r)} ~dv~d\omega~d\phi
\end{eqnarray*}
where $v = (-v_x\cos\phi + v_y\sin\phi)$.

This filtered back-projection formulation can be extended to the next dimension to calculate  $I(v_x,v_y,v_z)$ by changing $p(v,\phi)$ to $p(v,\phi,\psi)$, where $\psi$ is the orbital inclination, setting $v = (-v_x\cos\phi\sin\psi + v_y\sin\phi\sin\psi + v_z\sin\phi\cos\psi)$,  and integrating over $dv~d\omega~d\phi~d\psi$.   The 3D case reverts to the 2D case for $\psi = 90^\circ$, corresponding to total eclipses, compared to partial eclipses for lower inclinations.  Moreover, while orbital inclinations close to $90^\circ$ are preferred for 2D reconstructions, inclinations close to $45^\circ$ are optimal for 3D reconstructions \citep{richardsetal10}.

The application of Doppler tomography requires one main assumption, namely that the line profiles are broadened only by Doppler motions; which is a good first order approximation, although turbulent motions may contribute (see \citealt{kaitchucketal94} and the extensive discussion in \citealt{richardsetal10}).  In addition, we need numerous spectra distributed around the orbit of the binary, with high wavelength resolution and adequate coverage in orbital phase.  The technique has been applied to spectra that are dominated by emission, but the generality of the equations suggests that the method may also be applied with prudence to absorption spectra. 

In practice, tomography is a summation process and the velocity images are calculated by summing the normalized intensities of the spectra for all velocity pixels after they have been converted to the separate $v_x$ and $v_y$ components, and then sweeping through all phases around the orbit.  In the calculation, the Doppler shifts typically range from -800 to +800 {\kmps} relative to the rest wavelength of the line (e.g. H$\alpha$).  The images are constructed in the velocity domain and can only be converted to the Cartesian domain if the separate velocity fields of all accretion structures (e.g., gas stream, accretion disk, etc.) are known.

\begin{figure}[t]
\epsscale{1.2}
\hspace{-10pt}
\plotone{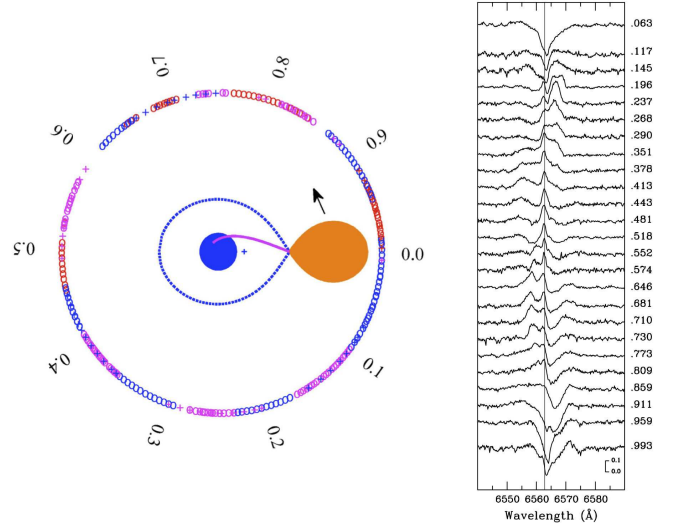}
 \caption{Full-orbit data collection over several epochs for U CrB  (left frame), and a representative set of these spectra stacked for epoch 1994 in order of orbital phase from 0.0 to 1.0 (right frame).   The arrow in the left frame shows the direction of orbital motion.  The spectra are summed in the tomography process relative to the rest wavelength of the H$\alpha$ line in the frame of reference of the mass-gaining star (vertical line).  
}
\label{fig3}
\end{figure}

Figure \ref{fig3} illustrates the full orbit data collection procedure in the case of U CrB, with 294 phases over several epochs.  A representative set of these spectra for epoch 1994 were stacked in order of orbital phase from 0.0 to 1.0 in preparation for the summation calculation.  Since the spectra of the direct-impact systems are usually dominated by absorption near primary eclipse ($\phi \sim 0.9 - 0.1$), these eclipse spectra were typically excluded from the analysis for these systems. The tomography calculations were implemented using both Fortran and C++ codes with MATLAB as the platform for creating the plots of the images.  The speed of the code depends sensitively on the number of orbital phases multiplied by the velocity resolution of the spectrum in pixels. 

\section{Application to Direct Impact Binaries}

Doppler tomography has been applied to eclipsing systems which contain white dwarfs and neutron stars (e.g., cataclysmic variables or CVs, nova-like systems, and soft X-ray binaries), as well as non-compact main sequence stars (e.g., Algol binaries).  The main differences between these systems are: (1) the accretion structures in the compact systems are bright relative to the stars, while the structures in the Algols are faint relative to the luminous main sequence mass-gaining star; and (2) the large size of the mass gainer in the Algols leads to the direct impact of the gas stream onto the stellar surface in the short-period Algols, while this type of impact does not occur in the long-period Algols or in compact systems.  Hence, composite accretion structures should form in the short-period Algols compared to the classical accretion disks that have been detected in other systems.  In all cases, the cool magnetically-active mass-losing star  is expected to influence the resulting accretion structures. 

Numerous time-resolved spectra of the target systems are required for the tomography calculation in order to recover a well-resolved image of the gas flows (see Figure \ref{fig3}).   Images are usually created from spectral lines in the optical regime since the expected gas temperatures are $\sim 10^4$K.  However, since the H$\alpha$ line is saturated in the spectra of CVs and X-ray binaries, relatively weaker lines have been used, e.g., He I $\lambda$5015, He II $\lambda$4686, H$\beta$. Many Doppler tomograms have been created for CVs (e.g., \citealt{marsh01,morales-rueda04,steeghs04,schwopeetal04}), X-ray binaries (e.g., \citealt{vrtileketal04} for SMC X-1 and Her X-1), and the nova-like binary V3885 Sgr \citep{prinjaetal11}.  The first tomogram of a black-hole candidate, Cyg X-1, was made by \citet{sharovaetal12}.

Most of the direct-impact Algol binaries have weak emission-line spectra, so the emission from non-photospheric gas can be enhanced by subtracting the composite spectrum of the stars from the observed spectrum to create a difference spectrum (e.g., \citealt{richards93}).  This subtraction process assumes that the circumstellar gas is optically thin, which is a good first order approximation.  However, this assumption can be examined by comparing the tomography images derived separately from the observed and difference spectra (see Section 6).   For the direct-impact systems, tomograms have been made predominantly from the H$\alpha$ line, and other images have been constructed from the H$\beta$, He I $\lambda$6678, and Si II $\lambda$6371 lines (e.g., CX Dra; \citealt{richardsetal00}), and also from the Si IV $\lambda$1063,1073 doublet and Si IV $\lambda$1394 line in the ultraviolet \citep{kempner+richards99}.  

\subsection{Evidence of Gravitational and Magnetic Activity}

Table \ref{tab3} provides a summary of spectra of Algol binaries (epochs 1976 to 2003) that were processed into images, and it lists the number of orbits, N(orbits), the number of phases, N(phi), and the velocity resolution of the spectra, N(vel).  These data  were derived primarily from the extensive database of direct-impact systems described by \citet{richards+albright99} for which the spectra have high signal-to-noise (typically S/N  $>  100$), high wavelength resolution (see Table \ref{tab3}), and coverage of the entire orbit of each binary (see Figure \ref{fig3}).   Other spectra were included for three systems:  $\beta$ Per (epochs 1976-1977; \citealt{richards92}),  TT Hya (epochs 1985-2001; \citealt{milleretal07}), and AU Mon (epochs 1984 - 2003; \citealt{atwood-stoneetal12}).  The best images are obtained from data with the highest number of phases distributed around the orbit (i.e., high N(phi)), and from spectra with the highest resolutions (i.e., high N(vel)).  Moreover, datasets collected over the smallest number of orbits are preferred since they show the state of the gas over fewer dynamical timescales (see Table \ref{tab2}).  

\begin{table}[t]
\caption{Data on a Variety of Algol Binaries and an RS CVn Binary}
\centering
\setlength{\tabcolsep}{4pt}
\begin{tabular}{llcccc}
\hline\hline \M
System & Epoch & N(nights) & N(orbits) & N(phi) & N(vel) \\
[0.5ex]\hline \M
RZ Cas	   &	1994 	&	~~7	&	~~~5.9	&	~~28	&	262	\\
$\delta$ Lib &	1993 	&	~~7	&	~~~3.0	&	~~51	&	275	\\
RW Tau	   &	1994 	&	~~7	&	~~~2.5	&	~~30	&	275	\\
$\beta$ Per &	1976-1977	&	23	&	156.9	&	~~59	&	728	\\
$\beta$ Per &	1992 &	~~7	&	~~~5.6	&	135	&	503	\\
$\beta$ Per &	1994 &	~~7	&	~~~2.4	&	~~36	&	275	\\
TX UMa   	   &	1992 &	15	&	~~~4.9	&	~~81	&	468	\\
TX UMa	   &	1993 &	29	&	~27.4	&	114	&	468	\\
TX UMa	   &	1994 &	14	&	~62.0	&	~~27	&	275	\\
U Sge	        &	1993 &	12	&	~~~9.5	&	106	&	256	\\
U Sge	        &	1994 &	~~7	&	~~~2.1	&	~~48	&	275	\\
S Equ	        &	1993 &	12	&	~~~9.3	&	~~57	&	275	\\
U CrB		  &	1993 	&	41	&	~24.3	&	161	&	459	\\
U CrB		  &	1994 	&	14	&	~55.0	&	~~47	&	275	\\
RS Vul	       &	1993 	&	12	&	~~~7.1	&	~~81	&	275	\\
SW Cyg	 &	1994	&	~~7	&	~~~1.5	&	~~36	&	275	\\
CX Dra	       &	1994	&	~~7	&	~~~1.0	&	~~54	&	275	\\
TT Hya	    &	1994-1997 &	 22 &	158.3	&	~~45	&	446	\\	
AU Mon   &  	1988-2003	&    64    &    458.2         &    ~~83       &    217     \\						
V711 Tau	 &	1994 	&	~~7	&	~~~2.5	&	~~27	&	209	\\
\hline
\end{tabular}
\label{tab3}
\end{table}

The 2D velocity images derived from back-projection Doppler tomography of the H$\alpha$ line are shown in four parts in Figure 4:  \ref{fig4a-f}, \ref{fig4g-l}, \ref{fig4m-r}, and \ref{fig4s-t}, for thirteen systems listed in order of orbital period from $P = 1.195$ days (RZ Cas) to $P = 11.113$ days (AU Mon).  In addition, the images of four systems are shown at multiple (2-3) epochs ($\beta$ Per, TX UMa, U Sge, U CrB).  These figures show the Cartesian representation of each binary, the trailed spectrogram, color and contour tomograms and the corresponding color bar for the images.   A reference template is overlaid on all images to show the expected locations of the stars, the gas stream gravitational trajectory, and the locus of a Keplerian accretion disk; these were illustrated in Figure \ref{fig2} and described in Section 2.   

In addition, since the mass-losing star in many interacting binaries is a cool magnetically-active star, we used the tomogram of the detached RS CVn binary V711 Tau (HR 1099) to illustrate the location of H$\alpha$ emission from a magnetically active star (see Table \ref{tab3}).  V711 Tau consists of a G star with a K star that nearly fills its Roche lobe; and the tomogram shown in Figure 4 (t) illustrates that most of the emission arises from the K-star component, which is similar to the mass loser in the direct-impact systems.

\setcounter{figure}{3} 
\renewcommand{\thefigure}{4(a-f)} 
\begin{figure*}[t]
\begin{center}
\epsscale{0.95}
\plotone{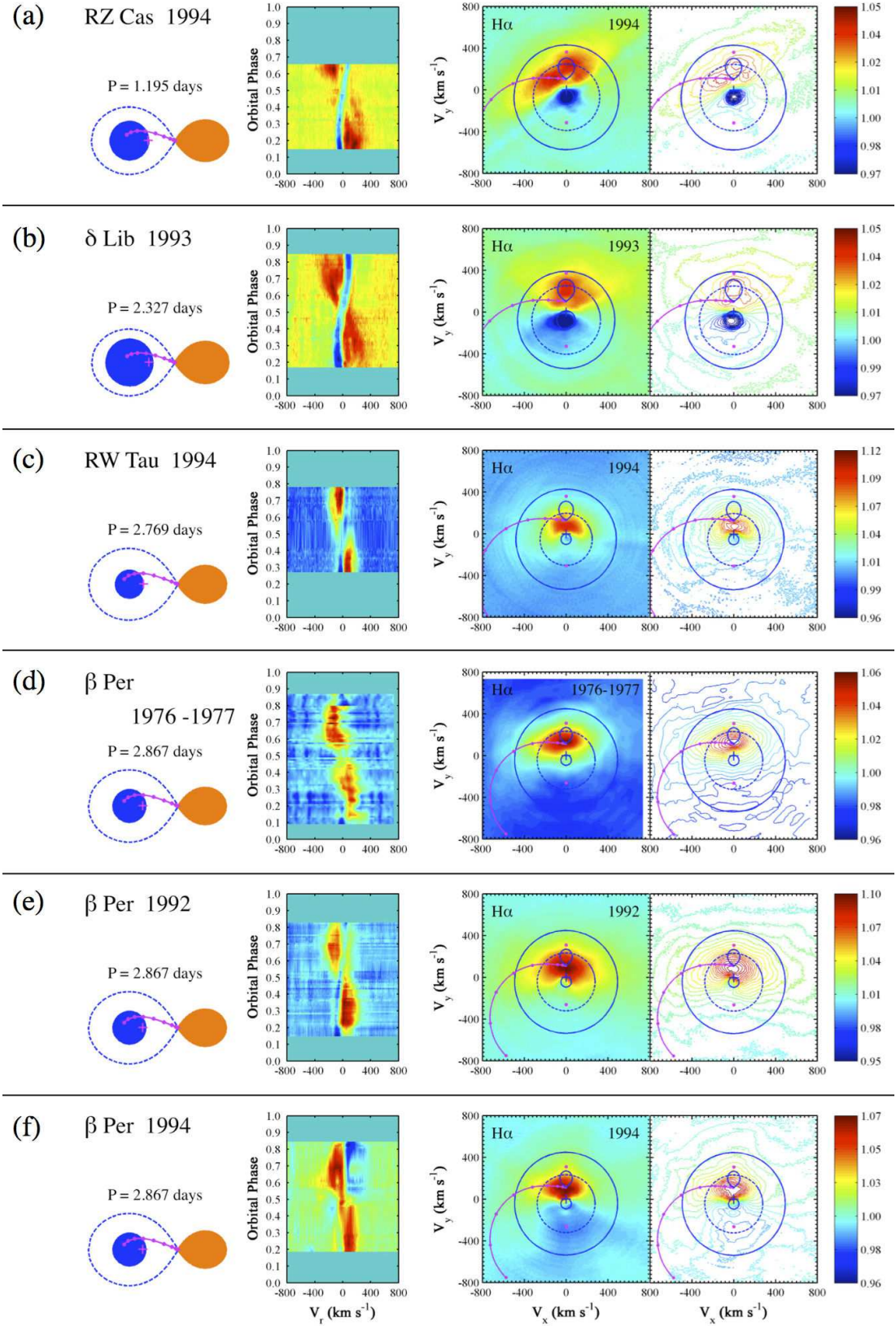}
\end{center}
 \caption{Two-dimensional H$\alpha$ Doppler tomograms of direct-impact and wider binaries in order of increasing orbital period: RZ Cas (epoch 1994),  $\delta$ Lib (epoch 1993), RW Tau (epoch 1994), $\beta$ Per (epochs 1976-77, 1992, 1994).  For each system, we show the Cartesian representation of the binary (left frame), with the trailed spectrogram (middle frame), and the color and contour tomograms (right frames), with color bar intensities in normalized units relative to the continuum.  The images were derived from difference spectra.}
 \label{fig4a-f}
\end{figure*}

\setcounter{figure}{3} 
\renewcommand{\thefigure}{4(g-l)} 
\begin{figure*}
\begin{center}
\epsscale{0.95}
\plotone{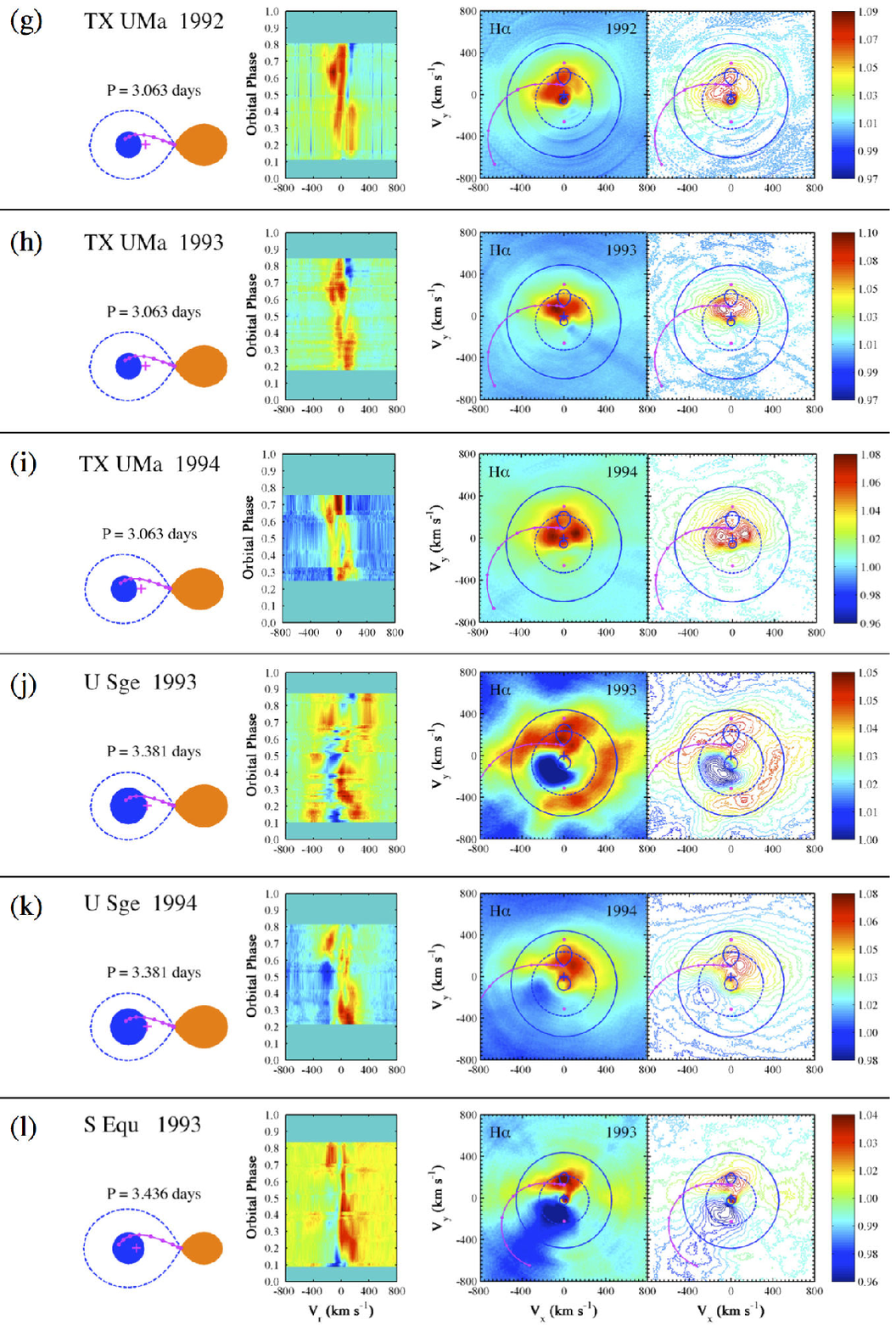}
\end{center}
 \caption{Same as Figure 4a for TX UMa (epochs 1992, 1993, 1994),  U Sge (epochs 1993, 1994), S Equ (epoch 1993). }
 \label{fig4g-l}
\end{figure*}

\setcounter{figure}{3} 
\renewcommand{\thefigure}{4(m-r)}
\begin{figure*}
\epsscale{0.95}
\begin{center}
\plotone{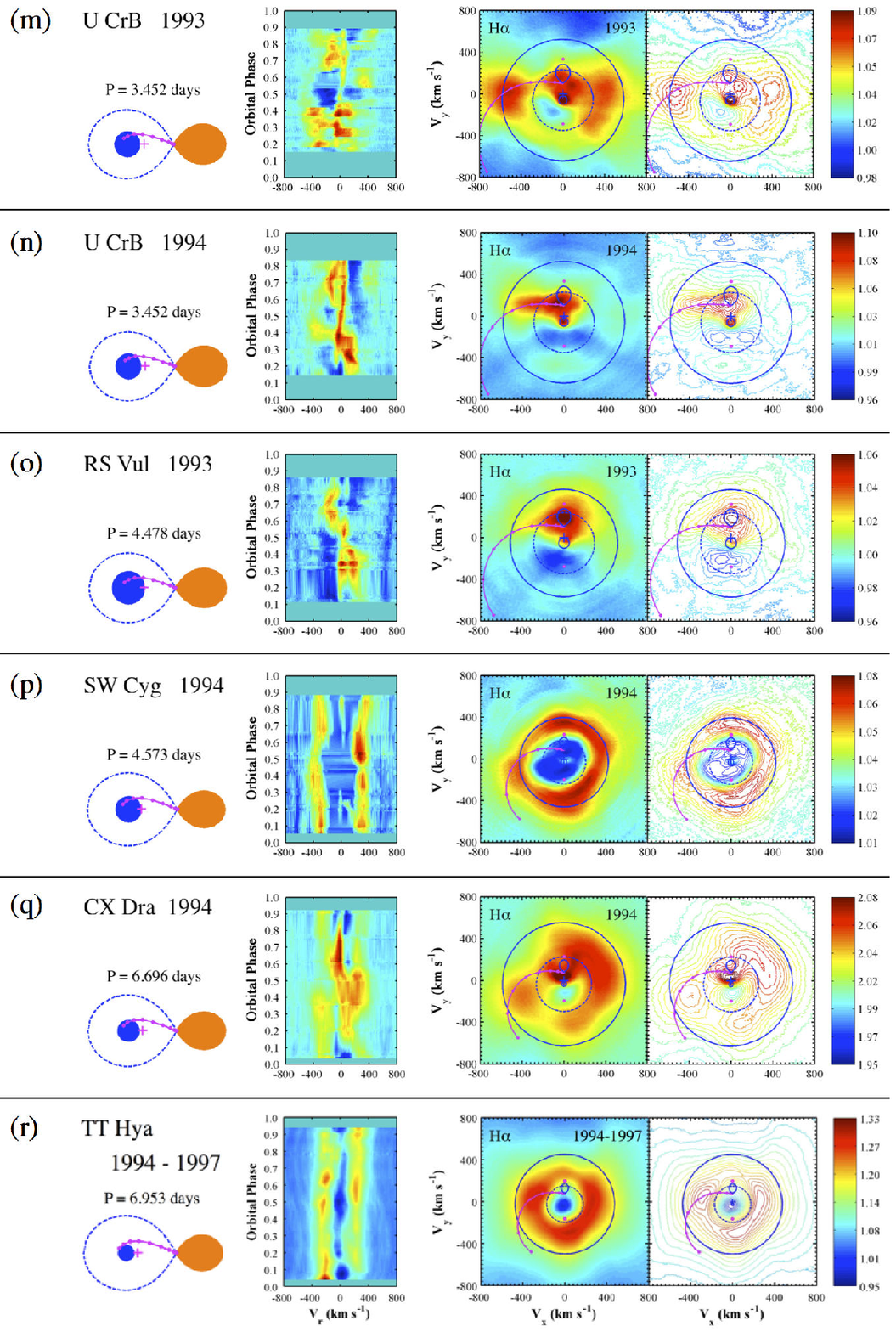}
\end{center}
 \caption{Same as Figure 4a for  U CrB (epochs 1993, 1994), RS Vul (epoch 1993), SW Cyg (epoch 1994), TT Hya (epochs 1994-1997). }
 \label{fig4m-r}
\end{figure*}

\setcounter{figure}{3} 
\renewcommand{\thefigure}{4(s-t)}
\begin{figure*}[t]
\epsscale{0.95}
\begin{center}
\plotone{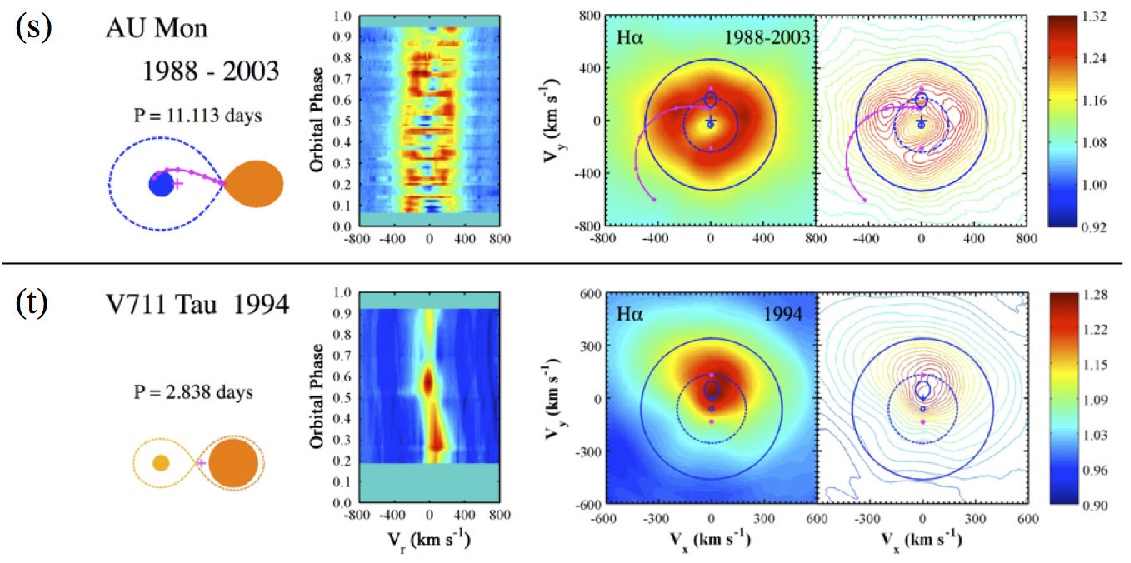}
\end{center}
\vspace{-10pt}
 \caption{Same as Figure 4a for  AU Mon (epochs 1988-2003) and the detached binary V711 Tau (epoch 1994), where the tomogram of the latter system was derived from observed spectra.  V711 Tau consists of a G star with a K star that nearly fills its Roche lobe, and the tomogram shows that most of the H$\alpha$ emission arises from the more active K star, which is similar to the mass loser in the direct-impact systems.}
 \label{fig4s-t}
\end{figure*}

The tomograms of Algol binaries in Figure 4 reveal several distinct sources of H$\alpha$ emission which result from a combination of gravitational forces and magnetic activity.  The main sources in the image are in accordance with the expected behavior of the gas flow from the L1 point under the influence of gravity and Coriolis forces, as described in Section 2.  In addition, nearly all systems display evidence of magnetic activity associated with the mass-losing star.  This latter result is surprising given that the mass-losing star contributes a small fraction of the total luminosity of the binary (only 5-15\%, depending on wavelength), and it was considered to be relatively unimportant in the mass transfer process, besides being the source of the gas flow through the L1 point.  

The sources of H$\alpha$ emission seen in the 2D images are described below:
\vskip5pt
\noindent
(1) {\it a source of magnetic activity associated with the cool mass-losing star}, including emission from the chromosphere, prominences, flares, and coronal mass ejections.  The tomogram of V711 Tau illustrated that the emission from the cool star has the same velocity as that star.  This type of emission was detected in almost all systems, including RZ Cas, $\delta$ Lib, $\beta$ Per, TX UMa, U Sge, S Equ, U CrB, RS Vul, CX Dra, and AU Mon; the exceptions were RW Tau, SW Cyg,  and TT Hya.  The byproducts of magnetic activity revealed in the velocity images of so many systems should not be surprising since the mass-losing stars have spectral types from F0 to K2 and there is independent evidence of magnetic activity in these systems \citep{richards+albright93,sarnaetal97}.

\vskip5pt
\noindent
(2) {\it a gas stream along the predicted gravitational trajectory} (e.g., RZ Cas, $\delta$ Lib, RW Tau, $\beta$ Per, TX UMa, U Sge, S Equ, U CrB, RS Vul, CX Dra, AU Mon).  In most of these systems, the gas stream flows along or close to the predicted gravitational trajectory.  The exception is TX UMa in which the stream path appears to be deflected during epochs 1992 and 1994 but not during epoch 1993.  It is interesting also that the emission associated with the cool star is weaker during epoch 1993 relative to the other epochs.  TX UMa is known to be magnetically active \citep{richards+albright93,sarnaetal98}, and therefore, it should not be surprising that enhanced magnetic activity in the form of prominences or coronal mass ejections could deflect the gas stream flow away from the expected gravitational trajectory.  

With increased room between the stars, the gas stream can make its way around the mass-gaining star to form a transient or stable accretion disk (e.g., TT Hya and AU Mon).  

\vskip5pt
\noindent
(3) {\it a star-stream impact region} where the gas stream strikes the surface of the mass-gaining star.  The expected location of this shock region was discussed in Section 2 and shown in Figure \ref{fig2} for different values of the size of the mass gainer.  In most cases, the shock should occur close to the intersection of the gas stream path and the circle representing the Keplerian velocity of the mass gainer.  Extended emission at this location is visible in the velocity images for U Sge (epoch 1993), U CrB (epoch 1993), and S Equ.

\vskip5pt
\noindent
(4) {\it a bulge of extended absorption or emission around the mass-gaining star} produced by the impact of the high velocity stream onto the slowly rotating photosphere (e.g., RZ Cas, $\delta$ Lib, TX UMa (epochs 1992 and 1994),  U Sge (both epochs), S Equ, U CrB (both epochs), and RS Vul).   This is the same region as the circumprimary bulge detected via photometry as a hot band around the equator of the mass gainer (e.g.,RZ Cas \citealt{olson82}; U Sge \citealt{olson87}).  The impact increases the rotational velocity of the star beyond the synchronous value \citep{etzel+olson93}.  Spectroscopic studies of TX UMa by \citet{komziketal08} and \citet{glazunovaetal11} also found that the mass-gainer in this binary rotates $\sim 1.5 - 2$ times the synchronous velocity as a result of the impact. So, the images have confirmed the photometric and spectroscopic evidence of  the circumprimary  bulge.

\vskip5pt
\noindent 
(5) {\it a transient or permanent accretion disk} around the mass-gaining star.  Accretion disks were found in all systems with $P_{orb} > 4.6^d$ (e.g., SW Cyg, CX Dra, TT Hya, AU Mon).  Disks were also found in U Sge (epoch 1994) and U CrB (epoch 1994), however the gas distribution in these two systems can be transformed from a disk-like state to a stream-like state within a few orbital cycles. The evidence of this variability can be seen in the trailed spectrograms and in the individual spectra through the change from a double-peaked profile to a single-peaked profile.  

\vskip5pt
\noindent
(6) {\it an absorption zone} where the gas flows are seen in absorption (e.g.,  U Sge, S Equ, U CrB, RS Vul, SW Cyg, CX Dra).    A tomogram created from the SiIV $\lambda$1394 line of U Sge is seen in Figure \ref{fig7} and shows that the main source of emission in the ultraviolet overlaps with the absorption zone, which in turn represents the locus of hotter gas ($T \sim 10^5$K). 

\vskip5pt
\noindent
(7) {\it a disk-stream impact region} where the circling gas strikes the incoming stream and is slowed by the impact (e.g.,  RW Tau, $\beta$ Per, TX UMa, U Sge, S Equ, U CrB, RS Vul, CX Dra).  The result is a region of emission nearly along the line of centers with velocities lower than that of a  Keplerian disk, i.e., within the locus of the disk in the velocity image.  Therefore, in the systems with $2.8^d \le P_{orb} < 6.7^d$, except for SW Cyg, after the gas stream strikes the star it has enough angular momentum to flow around the star until it comes in contact with the incoming stream.  In these systems, the density of the gas stream must be higher than the density of the gas that has circled the mass gainer.  This region has been called the Localized Region (e.g., $\beta$ Per, \citealt{richards92}; RW Tau, \citealt{vesper+honeycutt93}; other systems, \citealt{vesperetal01}). In wider binaries, the gas stream makes a more grazing impact and forms a more extensive flow in an accretion disk, so it is more difficult to determine where the disk intersects with the incoming stream in the velocity image.  

\vskip5pt
A pictorial representation of the variety of emission sources is shown in Figure \ref{fig5}.  This figure illustrates that the gas stream emission was sufficiently bright to be detected in the images of most systems.  

\setcounter{figure}{4} 
\renewcommand{\thefigure}{\arabic{figure}}
\begin{figure}[t]
\epsscale{1.1}
\plotone{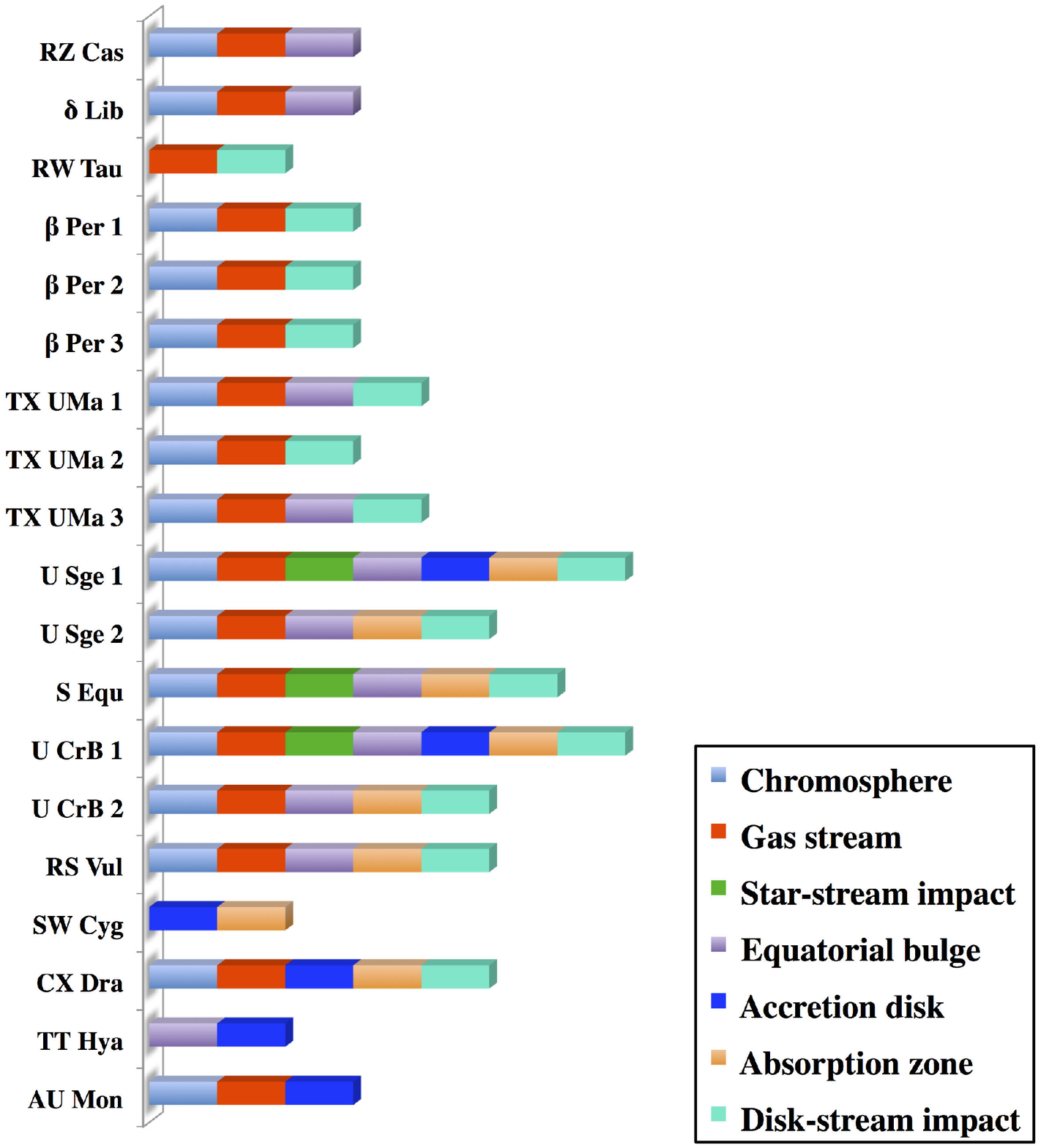}
 \caption{Types of emission structures identified in the tomograms, listed in order of increasing orbital period (top to bottom).}
 \label{fig5}
\end{figure}

\subsection{Transition from Stream-state to Disk-state}

\begin{figure}[t]
\epsscale{1.15}
\plotone{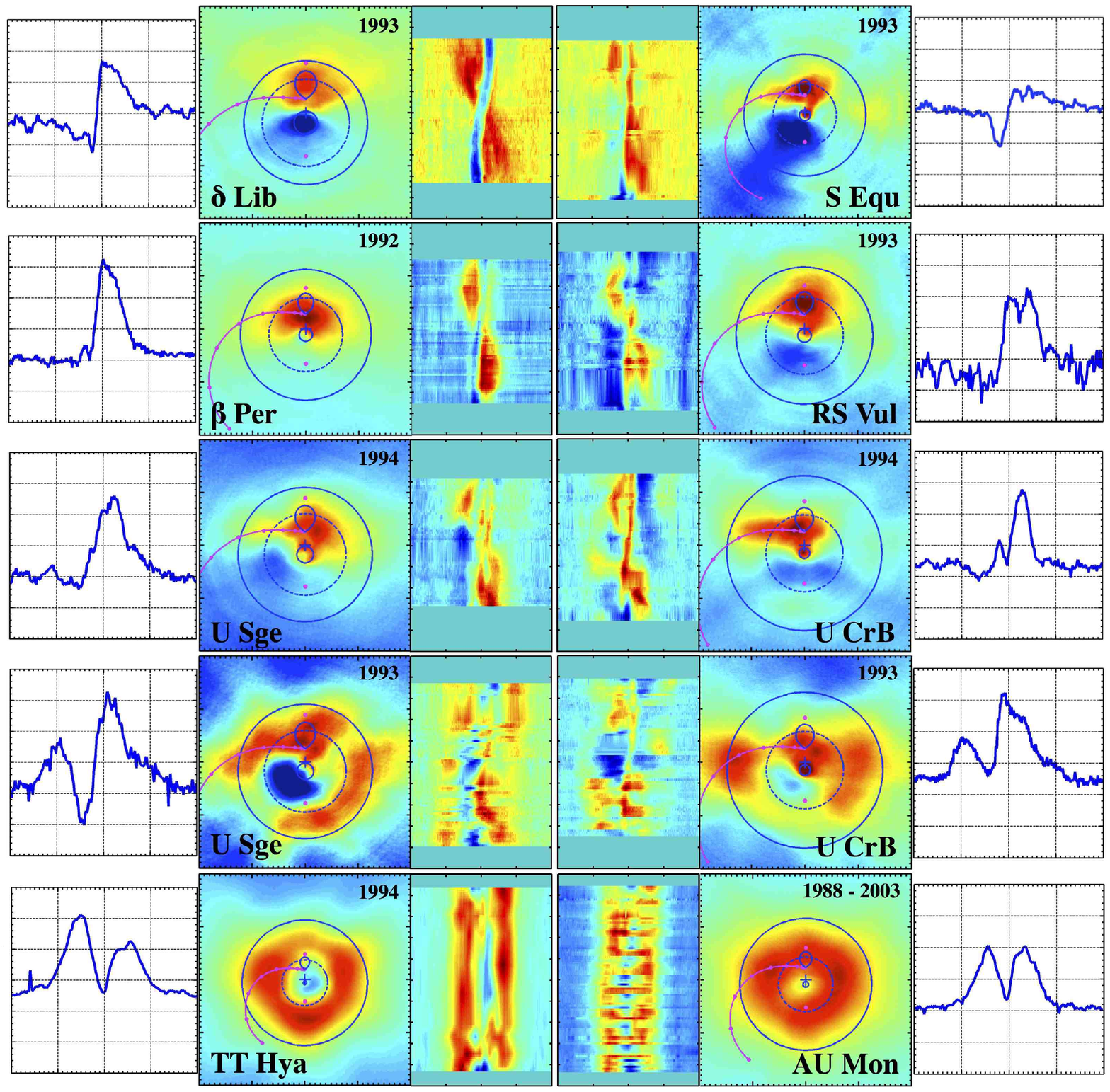}
 \caption{Summary of the transition in the line profile at quadrature, velocity image, and trailed spectrogram for a representative set of Algol binaries in our survey, from smaller binary separation (upper frames) to larger separation (lower frames). The progression illustrates the change from the stream state to the disk state.}
 \label{fig6}
\end{figure}

Figure \ref{fig6} summarizes the transition in the line profile at quadrature, the emission sources in the velocity image, and  the trailed spectrogram, with increasing binary separation (semi-major axis, $a$).  The progression from the upper to lower parts of this figure
illustrate the transformation from the {\it stream state} to the {\it disk state}.    In essence, the closest binaries (e.g., $\delta$ Lib) typically display only emission associated with the mass-losing star, and hence evidence of magnetic activity, because there is insufficient room for the gas stream to make its way around the mass gainer.  If there is some extra room between the stars, the gas stream can progress farther along, and we find evidence of extended gas streams along the projected gravitational path in several systems (e.g., S Equ, RS Vul, U Sge, U CrB).  In the widest systems (e.g., TT Hya, AU Mon), there is sufficient room to form a Keplerian accretion disk.  Over the range of systems, the line profile at quadrature shows a progression from nearly single-peaked profiles indicative of gas streams in the closest systems (smaller $a$), to distinct double-peaked profiles indicative of accretion disks in the wider systems (larger $a$).   

Potential variability in the images was examined in four systems for which data were collected at multiple epochs (i.e., $\beta$ Per, TX UMa, U CrB, U Sge), although data were also collected over an extensive range of epochs for CX Dra, TT Hya, and AU Mon.  The images of $\beta$ Per suggest that this system was relatively stable over many orbital cycles (epochs 1976 to 1994), however there was a slight change in the direction of the gas stream in TX UMa between epochs 1992 and 1994, suggesting the influence of magnetic activity on the cool star in that system (see earlier discussion in Section 4.1).  In addition,U CrB and U Sge are unusual because they display both stream-like and disk-like states at different epochs (see Figure \ref{fig6} and Section 4.1), and a closer examination of their spectra suggest that these binaries may alternate regularly between the two states.  Consequently, U CrB and U Sge have been designated as {\it Alternating Systems}, and the source of the variability may be linked to magnetic activity on the  mass-losing star (see \citealt{richards+waltmanetal03} and the discussion in Section 6).

The images of accretion disks in the wide grazing-impact systems are similar to those of the cataclysmic variables (e.g., \citealt{kaitchucketal94,szkodyetal00,szkodyetal01,prinjaetal11}).  Hence, stable accretion disks are found in binaries containing both compact and main sequence mass gainers.  

\subsection{Multiwavelength Observations}

\begin{figure}[t]
\epsscale{1.1}
\hspace{-10pt}
\plotone{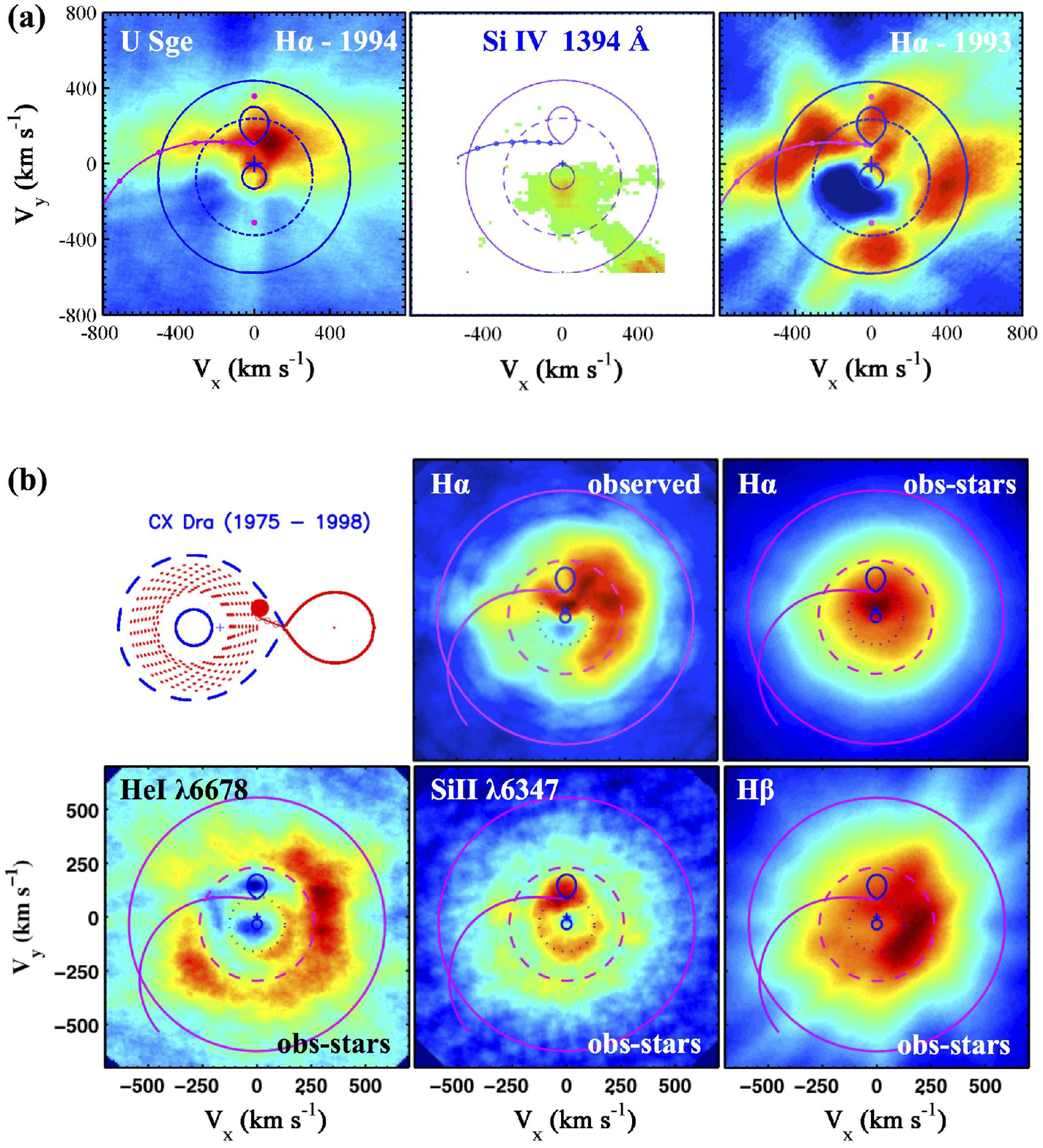} 
 \caption{Multiwavelength tomograms of (a) U Sge and (b) CX Dra showing the emission structures detected at different wavelengths.  The images of U Sge were obtained at H$\alpha$ (epochs 1993, 1994) and Si IV $\lambda$1394 (epoch 1983 - 1994; \citealt{kempner+richards99}), and for CX Dra at H$\alpha$, H$\beta$, He I $\lambda$6678, and Si II $\lambda$6371 (epochs 1975-1998; \citealt{richardsetal00}).   
 }
\label{fig7}
\end{figure}

Multiwavelength spectra of Algol binaries have been collected at wavelengths 4800 \AA ~$-$ 6700 \AA, including the $H\beta$, $H\alpha$, He I $\lambda$6678, and Si II $\lambda$6371 lines.  In addition, ultraviolet spectra of 17 Algol binaries can be found in the IUE archives from epochs 1978 to 1996.  The first ultraviolet image of U Sge was derived from the Si IV $\lambda$1394 doublet (epoch 1983-1994) after the contribution of the stars was extracted from the spectra. Figure \ref{fig7}(a) shows that the Si IV emission overlaps with the absorption zone identified in the H$\alpha$ image for epoch 1993, and can be interpreted as the locus of gas that was heated beyond the $10^4$K regime associated with H$\alpha$ to temperatures of $10^5$K in the UV region \citep{kempner+richards99}.  Figure \ref{fig7}(b) displays the multiwavelength images of CX Dra, and demonstrates that the He I $\lambda$6678 and Si II $\lambda$6371 emission arise from an extended accretion disk, while the H$\alpha$ and H$\beta$ emission arise primarily from a compact region where the gas stream strikes the outer edge of the disk \citep{richardsetal00}.
These results suggest that a more comprehensive understanding of the gas flows can be obtained when images are constructed for different temperature regimes.

\section{Influence of the System Properties}

\begin{figure*}[t]
\epsscale{1.15}
\hspace{-10pt}
\plotone{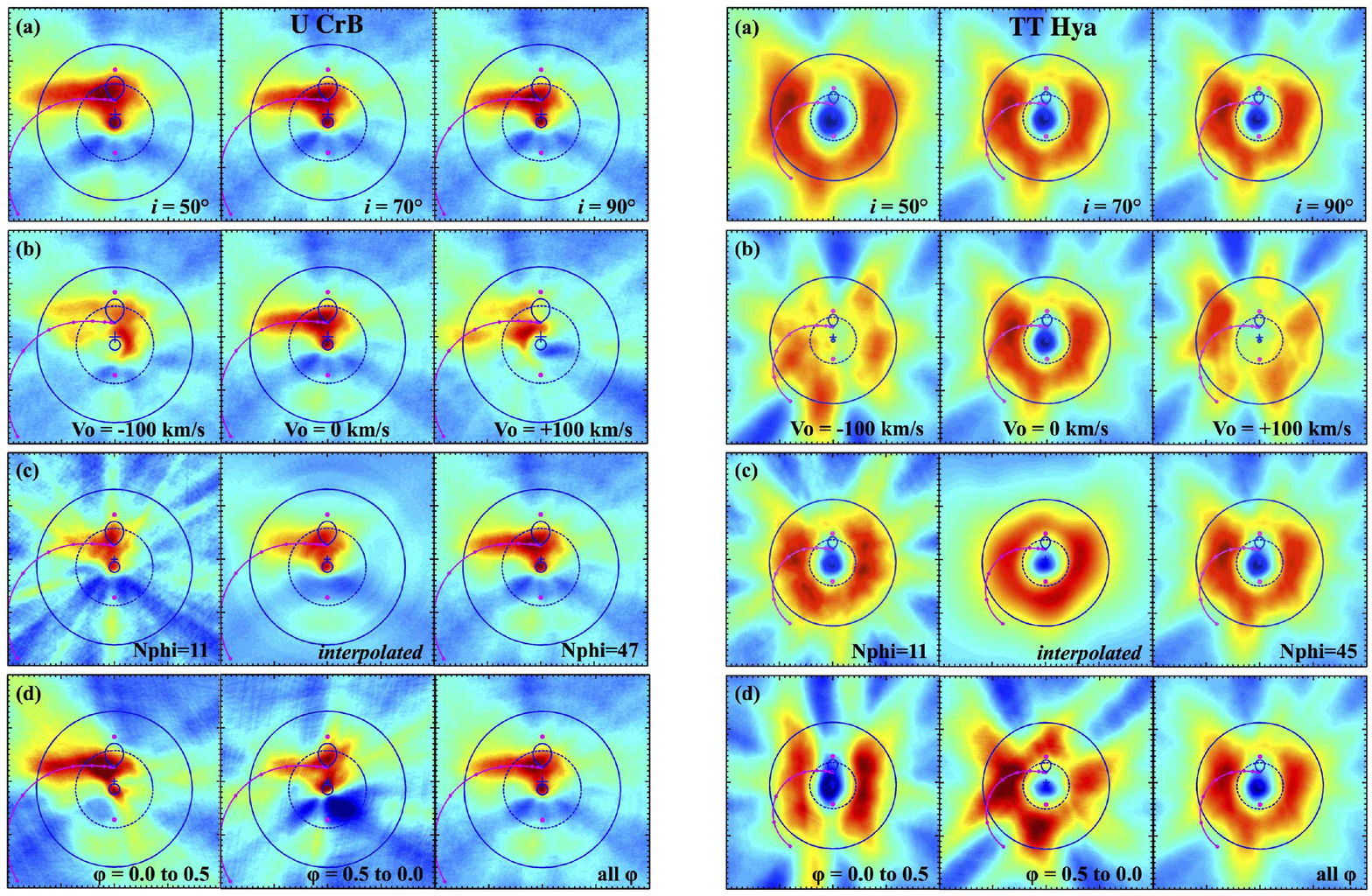} 
 \caption{Effects of the parameters on the 2D image: (a) orbital inclination: $i = 50^\circ - 90^\circ$; (b) systemic velocity: $V_o = -100, 0, +100$ {\kmps}; (c) phase coverage around the orbit for a subset of the data,  an interpolation of the subset using 100 evenly-spaced phases, and the full dataset; and (d) shadowing study using data at phases $\phi=0.0 - 0.5$, $\phi=0.5 - 1.0$, and the complete data set.  The results are shown for the cases of U CrB (epoch 1994; left frames) and TT Hya (epoch 1994-1997; right frames).
The measured inclinations and systemic velocities of U CrB and TT Hya are ($79.1^\circ$, $-6.7$ {\kmps}) and ($84.4^\circ$, $0.0$ {\kmps}), respectively.   }
\label{fig8}
\end{figure*}

The attractiveness of tomography is that it is a robust technique; however it is useful to discover how various criteria influence the resulting 2D image constructed from the back-projection technique.  It should be noted that the tomograms shown in Figure 4 were computed without any filtering, smoothing, denoising or other regularization of the images. The only additional processing was to interpolate the images in orbital phase to reduce the radial spoke-like streaks produced by the back-projection procedure when there are gaps in orbital phase.  In spite of the minimal processing, the amount of detail manifested in the images is truly remarkable. 

In this section, we examine the effects of (1) orbital inclination, (2) systemic velocity, (3) orbital phase coverage, (4) velocity resolution, and (5) shadowing on the tomography results.  

\subsection{Orbital Inclination}
The effects of adopting different values of orbital inclination are shown in Figure \ref{fig8} for a direct-impact binary (e.g., U CrB)  and a wider binary  (e.g., TT Hya).  The main result is that the image size decreases as the inclination increases from $50^\circ$ to $90^\circ$. Moreover, the measured inclinations of U CrB ($i = 79.1^\circ$) and TT Hya ($i = 84.4^\circ$) lead to images that are consistent with the template of predicted locations for the gas stream and disk in the velocity frame. 

\subsection{Systemic Velocity}
Figure \ref{fig8} illustrates the effect of adopting a systemic velocity for the binary, $V_o$, that is inconsistent with the measured value; the result is a distortion of the image.    The example cases of U CrB and TT Hya have systemic velocities of $-6.7$ {\kmps} and $0.0$ {\kmps}, respectively, and the corresponding 2D images are therefore similar to the image for $V_o = 0$ {\kmps}.  

\subsection{Orbital Phase Coverage and Velocity Resolution}
Figure \ref{fig8} illustrates the effect of phase coverage on the resulting 2D image using two subsets the U CrB and TT Hya data.  
The images were made from three sets of data:  a subset of the full dataset (left frame), after this subset was interpolated to create a grid of 100 evenly-spaced phases (middle frame), and the full dataset (right frame).  Insufficient orbital phase coverage creates radial spoke-like streaks across the diagram.  However, an interpolation in orbital phase adequately represents the outcome when the full dataset is used to construct the image.

The resolution of the H$\alpha$ line was typically 0.17 \AA/pixel (N(vel) = 275) and was as fine as 0.09 \AA/pixel (N(vel) = 503).  Since the wavelength or velocity resolution of the spectra is already very high, an increase in the resolution does not result in any significant improvement in the resulting 2D image. However, poor wavelength resolution in the spectrum would lead to a coarser image.

\subsection{Shadowing and the Transparency of the Medium}

The effect of shadowing results when the passage of light through a medium is obstructed because of the opacity of the medium.  Hence, the view from one direction may not be identical to that from the opposite direction $180^\circ$ away.  A similar effect called an acoustic shadow in radar tomography occurs when radio waves fail to propagate due to some topological obstruction, or in medical tomography because of the attenuation of light by denser parts of the medium \citep{girardetal11}.  

Figure \ref{fig8} shows images produced from data at phases $\phi=0.0 - 0.5$ (left), $\phi=0.5 - 1.0$ (middle), and the complete data set (right) to check for the effect of shadowing in the example cases of U CrB and TT Hya.   The rough resemblance between the images derived from half the orbit and that from the full orbit suggest that shadowing, if it exists, does not play a significant factor in the resulting tomograms.  Instead, the uneven sampling of the data for each half orbit and the number of spectra in each dataset have larger effects on the resulting image.

An alternative way to examine the opacity of the medium is to compare the tomograms derived from the observed and difference spectra, since the subtraction of the stellar spectra from the observed spectrum assumes that the circumstellar gas is optically thin.  In the case of CX Dra, the observed and difference H$\alpha$ images shown in Figure \ref{fig7}(b) are not similar in appearance, which suggests that the circumstellar gas in CX Dra was not optically thin.  However, there is evidence from similar comparisons that the circumstellar gas is optically thin in TT Hya and AU Mon (see Section 6.3), in agreement with the results of the shadowing test for TT Hya (Figure \ref{fig8}).

\section{Physical Properties derived from the Images}

The 2D velocity images have provided visual confirmation of the existence of both gravitational and magnetic structures in the direct-impact and grazing-impact Algol binaries.  In addition, the physical properties of the gas flows can be derived from the data and images using (1) an analysis of the S-wave patterns in the trailed spectrograms, (2) the creation of simulated tomograms derived from hydrodynamic simulations, (3) the computation of synthetic spectra of the accretion structures, and (4) a systematic extraction of the separate accretion structures using synthetic spectra with tomography.   These procedures require substantial effort and we present here some examples of the results that have been achieved so far.  These results have enhanced our interpretation and understanding of the gas flows in both the velocity and Cartesian frames.

\vskip15pt
\subsection{S-wave Analysis}

\begin{figure}[t]
\begin{center}
\epsscale{1.1}
\plotone{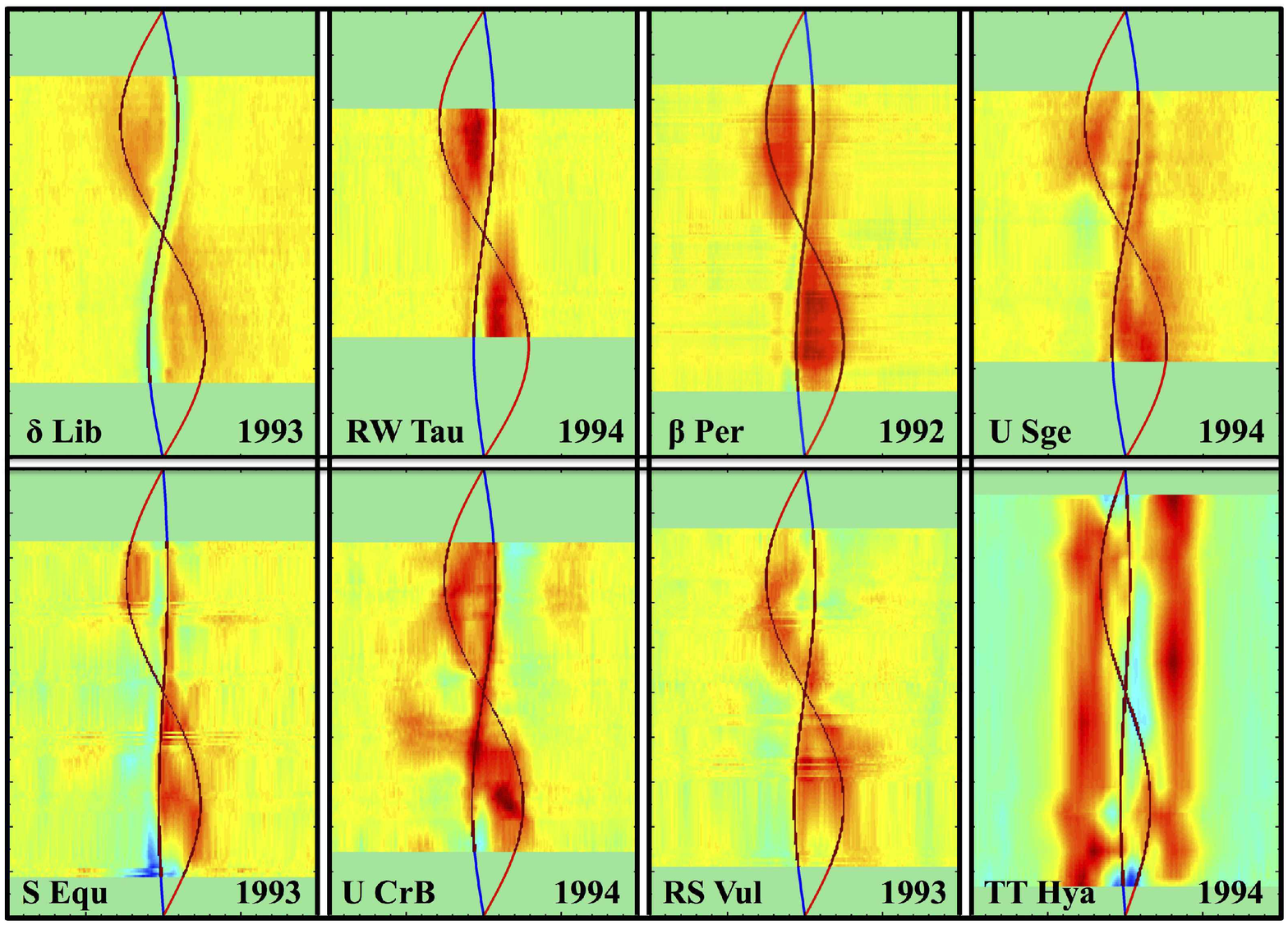} 
\end{center}
\vspace{-0.2cm}
 \caption{Velocity curves of the stars superimposed on the  trailed spectra.  The first seven frames show that the H$\alpha$ emission peaks roughly follow the velocity curves of the stars in direct-impact systems when they are in the stream state (e.g., $\delta$ Lib, RW Tau, $\beta$ Per, U Sge, S Equ, U CrB, RS Vul).  However, the emission peaks appear to be independent of the velocity curves in last frame (bottom right) when the gas is in a disk state (e.g., TT Hya).}
 \label{fig9}
\end{figure}

The trailed spectra of the direct-impact binaries display one or two sinusoidal patterns called {\it S-waves} which represent the orbital  radial velocity variations of the emission peaks (see Figures 4 and \ref{fig6}).  These patterns can be used to identify the locations of the most intense sources of emission in the velocity image.  Figure \ref{fig9} shows that the H$\alpha$ emission peaks roughly follow the velocity curves of the stars, sometimes with a smaller amplitude, in direct-impact systems when they are in the stream state (e.g., $\delta$ Lib, RW Tau, $\beta$ Per, U Sge, S Equ, U CrB, RS Vul).  However, the emission peaks appear to be independent of the velocity curves when the gas is in a disk state (e.g., TT Hya).   


In the direct-impact binaries, the larger variation is associated with the mass-losing star so the ratio of velocity semi-amplitudes of the mass loser and the S-wave source should be the same as the ratio of their distances from the center of mass.  As a result, this S-wave source is located at a distance $a_s =  a_2 \times (K_s/K_2)$ from the c.m. of the binary, where $K_2$ is the velocity semi-amplitude of the mass loser, $a_2 =  a/(1 +q)$ is the distance of the mass loser from the c.m. of the binary, and $K_s$ is the velocity semi-amplitude of the S-wave.   A similar argument follows for the other S-wave associated with the mass-gaining star.
 
\begin{figure}[t]
\begin{center}
\epsscale{1.1}
\plotone{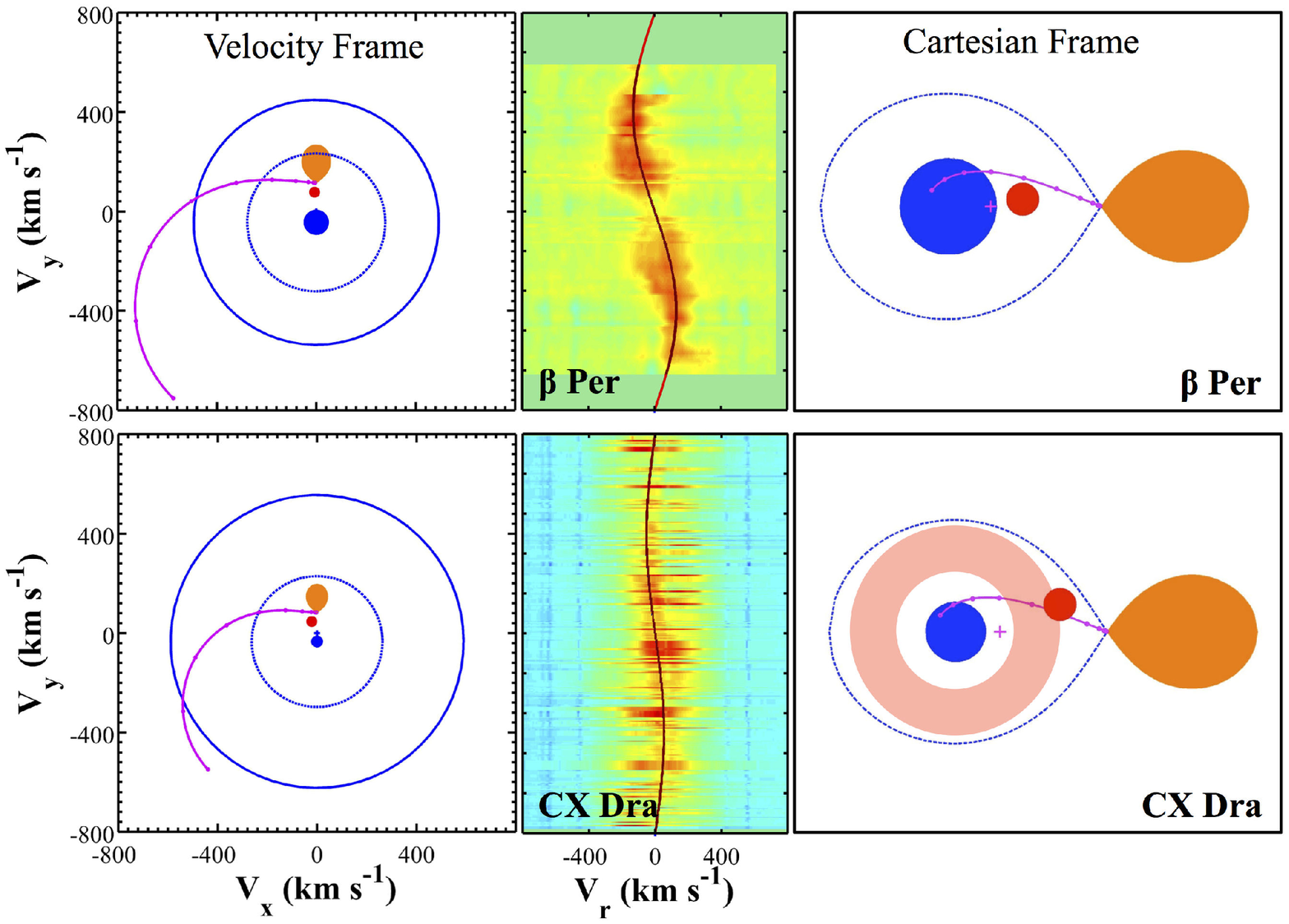} 
\end{center}
\vspace{-0.4cm}
 \caption{Illustration of the S-wave results for $\beta$ Per (1976-1977; upper frames) and CX Dra (1975-1998; lower frames).  The dominant S-wave solution in the velocity frame (large red dot; left frame), the S-wave superimposed on the trailed spectrum (middle frame), and the predicted source of the dominant S-wave in the Cartesian frame (large red dot; right frame) are shown along with the predicted locus of the accretion disk in CX Dra.}
 \label{fig10}
\end{figure}

S-wave analyses for $\beta$ Per (epochs 1976-1977) and CX Dra (epochs 1975 - 1998) have identified the location of the most intense sources of H$\alpha$ emission in these binaries.  Figure \ref{fig10} illustrates the dominant S-wave solution in the velocity frame (large red dot; left frame), the S-wave superimposed on the trailed spectrum (middle frame), and the predicted source of the dominant S-wave in the Cartesian frame (large red dot; right frame) for these binaries.  

For the case of $\beta$ Per (epoch 1976-1977), the source of the dominant S-wave was located nearly along the line of centers between the stars, with coordinates $(V_x,V_y) = (-7.5, 77.3)$ {\kmps} relative to the center of mass of the binary \citep{richards+jones+swain96}.  The corresponding source in Cartesian coordinates was found at a distance $a_s =  0.53a$ compared to the location of the $L1$ point at $0.650a$ from the c.m. of the binary, for $K_2 = 201$ {\kmps} and $q = 0.22$. This source is consistent with the location of the Localized Region in the Cartesian map predicted by \citet{richards93}.   

The analysis of CX Dra (epochs 1975-1998) identified the dominant S-wave source at $(V_x,V_y) = (-21.0, 46.0)$ {\kmps} relative to the c.m. of the binary \citep{richardsetal00}. The derived value of $K_s$ = $52 \pm 3$ {\kmps} led to $a_s = 0.30a$ from the c.m. of the binary compared to $a_{L1}$ =  $0.645a$, for $K_2 = 146$ {\kmps} and $q = 0.23$.  Figure \ref{fig10} shows the position of the S-wave source relative to the predicted locus of the accretion disk derived from the HeI $\lambda$6678 line (shown in Figure \ref{fig7}). This S-wave solution is consistent with the H$\alpha$ velocity image and suggests that the H$\alpha$ emission along the gas stream path was interrupted by the presence of a slightly hotter accretion disk.  Hence, this source represents the location of the {\it stream-disk impact site or ``hot spot''} and should be added to the list of seven emission sources described in Section 4.1.  Similar analyses can be performed for other systems. 

\subsection{Simulated Velocity Tomograms Derived from Hydrodynamic Calculations}

Hydrodynamic simulations are useful in the study of accretion flows because they illustrate the evolution of the flow under a variety of initial conditions, and can be used to constrain the properties of the gas flows derived from spectroscopic analysis.  These simulations can be used to model the H$\alpha$ emissivity and to study the evolution of the flow simultaneously in spatial and velocity dimensions. This was achieved by \citet{richards+ratliff98} for two Algol binaries: $\beta$ Per and TT Hya, which span the range from direct-impact to disk-state systems.  They used the VH1 (Virginia Hydrodynamics-1) 2D hydrodynamics code, which incorporates the effects of gravity and Coriolis forces, and assumes optically-thin radiative cooling of the gas; no magnetic effects were included in these simulations.  

\begin{figure}[t]
\begin{center}
\epsscale{1.1}
\plotone{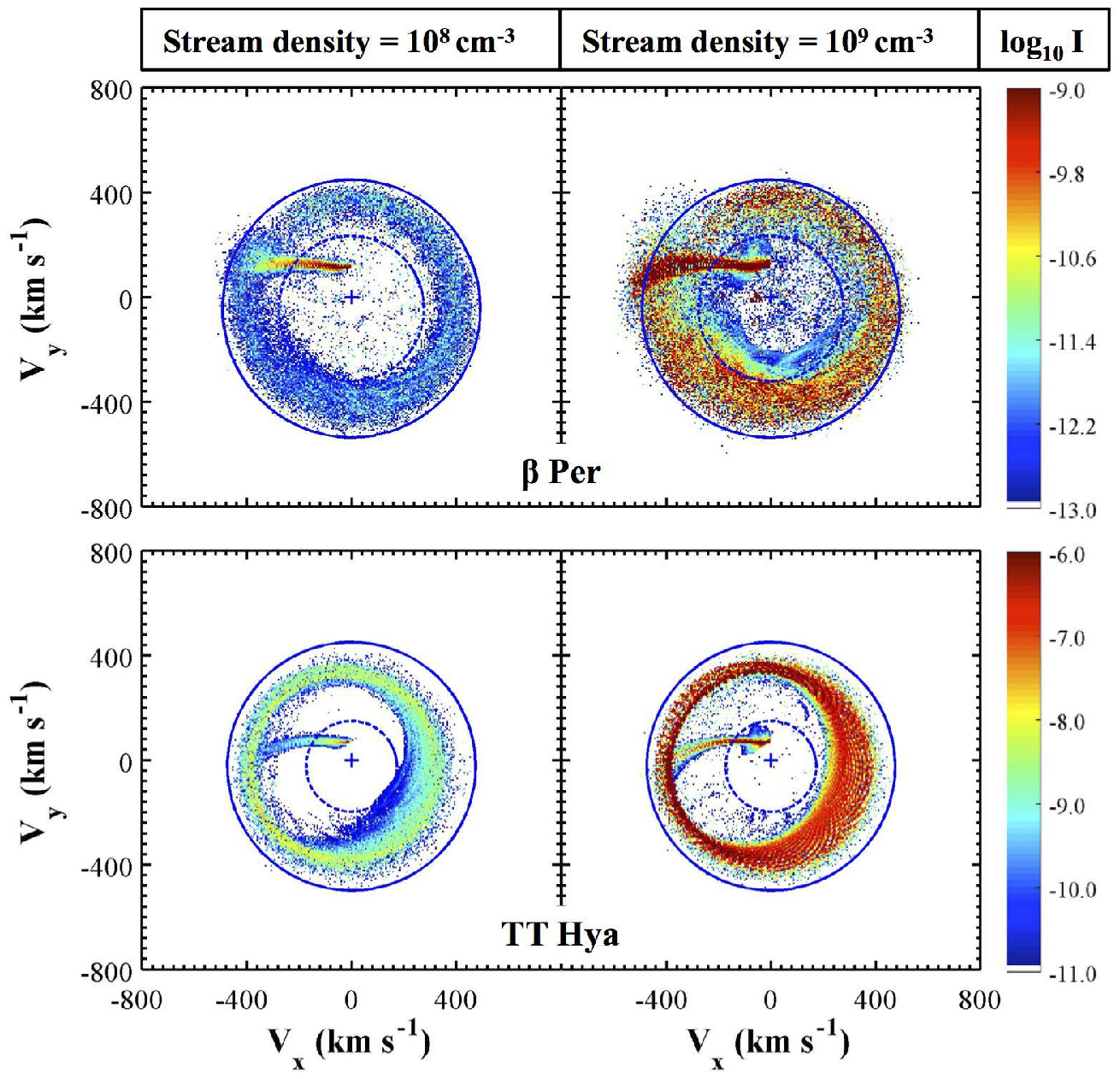} 
\end{center}
\vspace{-10pt}
\caption{Simulated velocity tomograms of $\beta$ Per (top frames) and TT Hya (bottom frames) based on hydrodynamic simulations for two values of the initial density of the gas stream at the L1 point: $10^8$ cm$^{-3}$ (left frames) and $10^9$ cm$^{-3}$ (right frames).  The H$\alpha$ emissivity, I, is given in powers of 10 on the color bar.}
 \label{fig11}
\end{figure}

For each binary, a {\it simulated velocity tomogram} can be constructed by summing the velocity images derived from the simulations over a range of epochs, $t = (2 - 4) P$, where $P$ is the orbital period, for comparison with the Doppler tomograms computed from observed spectra.  Figure \ref{fig11} shows the simulated velocity tomograms of $\beta$ Per and TT Hya for two values of the initial density of the gas stream at the L1 point: $10^8$ cm$^{-3}$ and $10^9$ cm$^{-3}$. In the case of $\beta$ Per, only the gas stream would be observed at the lower density because the remaining gas is nearly a factor of $10^3$ fainter than the gas stream, while both the gas stream and the disk would be observed at the higher density because they have comparable intensities.  Similarly, the gas stream and disk are substantially brighter in TT Hya at the higher density.   

The similarity between the H$\alpha$ Doppler tomograms of the alternating systems at different epochs (specifically U CrB and U Sge) and the simulated tomograms of $\beta$ Per at the two densities suggest that the variable and alternating behavior of the gas flows may be associated with changes in the density of the gas stream, and hence with changes in the mass transfer rate \citep{richards+ratliff98}.   Moreover, since the change between stream state and disk state in U CrB and U Sge can occur within days, which is much shorter than the Kelvin-Helmholtz timescales listed in Table \ref{tab2}, \citet{richards+waltmanetal03} proposed that the alternating behavior may be triggered by flaring activity on the mass-losing star and could be associated with magnetic activity on the cool star.   

The significant flaring cycles for $\beta$ Per (a direct-impact system) and for V711 Tau and UX Ari (two RS CVn binaries) at radio frequencies of 2.3 GHz and 8.3 GHz are given in Table \ref{tab4} based on a 5.6-year nearly-continuous survey of radio flares \citep{richards+waltmanetal03}.  These results suggest that flares could influence the mass transfer process regularly, as often as every 8 - 42 orbital cycles.  This survey demonstrated that radio flares from $\beta$ Per occurred with a predictable cycle of 49 days, which could  result in sporadic mass ejections through the L1 point on that timescale.

An alternative explanation is that the change in the image could result from the interaction between the incoming gas stream and the gas that has circled the star since the variability timescale of days is comparable to the circularization timescale $t_{circ}$ (see Table \ref{tab2}).

In the case of TT Hya, the hydrodynamic simulations produced an accretion disk that was asymmetric in both Cartesian space and velocity coordinates (Figure \ref{fig11}). This important result  went unnoticed for nearly 10 years until a synthetic spectrum of a symmetric accretion disk was modeled, extracted from the observed spectrum, and used to construct a difference Doppler tomogram of TT Hya \citep{milleretal07}.  This procedure revealed a small arc of emission within the locus of the accretion disk in the Doppler tomogram exactly at the location of the disk asymmetry seen in the simulated tomogram (see Figure \ref{fig12} and Section 6.3).  In summary, the construction of simulated tomograms from hydrodynamic calculations under controlled conditions have improved our understanding of the geometry and dynamics of the accretion flows in both velocity and Cartesian dimensions.

\begin{table}[t]
\caption{Periodicities of Radio Flares at 2.3 GHz and 8.3 GHz}
\centering
\setlength{\tabcolsep}{4pt}
\begin{tabular}{lcll}
\hline\hline \M
System & $P_{orb}$ (d) & Flaring cycles, $P$ & $P/P_{orb}$ \\
[0.5ex]\hline \M
$\beta$ Per & 2.87& 48.9 $\pm$ 1.7	& 17.1 \\					
V711 Tau	 &	2.84 & 120.7 $\pm$ 3.4, 80.8 $\pm$ 2.5 & 42.5, 28.5	\\
UX Ari & 2.33 & 141.4 $\pm$ 4.5, 52.6 $\pm$ 0.7 & 22.0, 8.2 \\
\hline
\end{tabular}
\label{tab4}
\end{table}

\subsection{Extraction of Emission Sources using Synthetic Spectra}

The computation of synthetic spectra provides us with a tool that can be used to extract the separate sources of emission from the composite tomograms of the direct-impact and disk-state Algol systems.   The {\sc shellspec} code can create composite spectra and light curves of interacting binaries which contain moving transparent or non-transparent 3D structures such as a disk, stream, spot, jet, or shell, in addition to the stars.  This code solves the simple radiative transfer along the line of sight in LTE, in an optional optically thin 3D moving medium (see \citealt{budaj+richards04} for details).  In addition, \citet{budajetal05} used the code to examine the effects of various free parameters and how double-peaked emission from an accretion disk is formed.  They found that the overall strength of the emission is influenced mainly by the density and temperature of the disk, while the position and separation of the emission peaks are influenced by the outer disk radius, the inclination, and the radial density profile. Furthermore, the depth of the central absorption in the profile is very sensitive to the temperature, inclination, geometry, and dynamics of the disk.  

The {\sc shellspec} code was used to sequentially model the composite synthetic spectra of the (i) stars, (ii) stars and disk, (iii)  stars, disk, and gas stream, or (iv) stars, disk, gas stream, and spot (e.g., Localized Region).  These calculations require the availability of optimal values for the orbital elements and stellar properties, and well as initial estimates of the temperature, mass density and electron number density of the disk. The estimated density of the stream at the L1 point was chosen to improve the fit between the observed and synthetic emission at the quadratures of the orbit and during the eclipses.  The density was allowed to vary along the stream to satisfy the continuity equation, and the electron number densities in the disk and stream were calculated from the temperature, density, and chemical composition using the Newton-Raphson linearization method.  The resulting composite synthetic spectrum provides a good comparison with  the observed H$\alpha$ spectra for TT Hya \citep{milleretal07}  and AU Mon \citep{atwood-stoneetal12}.

\begin{figure*}[t]
\begin{center}
\epsscale{1.1}
\plotone{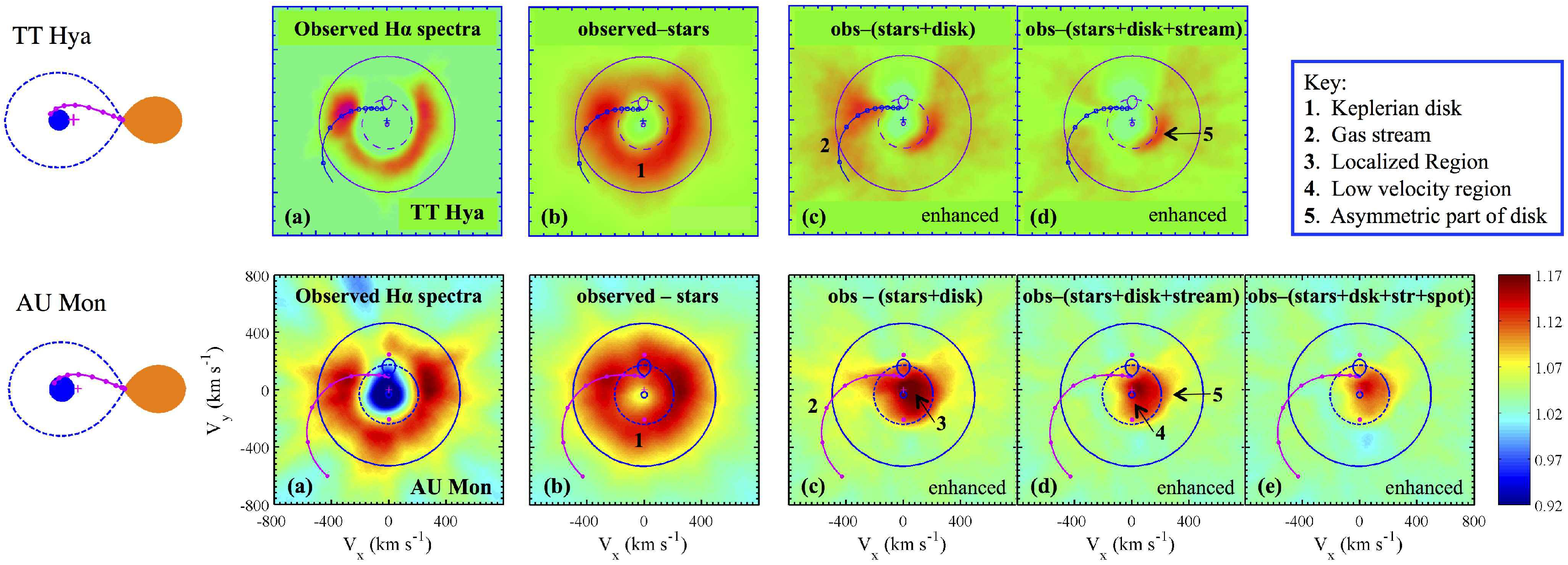} 
\end{center}
 \vspace{-10pt}
 \caption{Two-dimensional tomograms of the disk-state systems TT Hya (upper frames) and AU Mon (lower frames) illustrating the use of {\sc shellspec} modeling to isolate the separate accretion structures (disk, gas stream, Localized Region).  The images are derived from the (a) observed spectra, and when various accretion structures are subtracted from the observed spectra: (b) stellar spectra removed, (c) stars and disk removed, (d) stars, disk and stream removed, and (e) stars, disk, stream and Localized Region removed (AU Mon only).  }
\label{fig12}
\end{figure*}

The comparison between the observed and synthetic spectra can provide us with the physical properties of the accretion structures.  For TT Hya, \citet{milleretal07} adopted an initial disk temperature of $7000$ K and found that the disk extends up to about 10 solar radii and has a vertical thickness comparable to the diameter of the primary star, with mass density of $\sim$$10^{-14}$ g\,cm$^{-3}$ and electron number density of $\sim$$10^{10}$ cm$^{-3}$.  The initial gas stream temperature was set to $8000$ K, corresponding to the peak in H$\alpha$ emissivity, and the density of the gas stream at the $L_1$ point was estimated to be $2\times10^{-14}$g\,cm$^{-3}$. Finally, a mass transfer rate of $\sim2\times10^{-10}M_{\odot}yr^{-1}$ was derived from the velocity, density, and cross-section of the gas stream above and below the disk.   A similar analysis for AU Mon revealed a mass transfer rate of  $\sim2.4\times10^{-9}M_{\odot}yr^{-1}$ \citep{atwood-stoneetal12}.

The synthetic spectra can also be used in combination with tomography to demonstrate visually that the separate sources of emission have been adequately modeled. This procedure has been accomplished for two disk-state systems, TT Hya and AU Mon, and the results are displayed in Figure \ref{fig12}. The refined properties derived from the comparison between the observed and synthetic spectra were used to generate composite synthetic spectra of the disk, stream, and other features for input to the tomography code.  

The sequential modeling and extraction of the spectral contributions of the stars and various accretion structures resulted in the identification of an asymmetric accretion disk in both TT Hya  and AU Mon \citep{milleretal07,atwood-stoneetal12}.   The residual emission in the TT Hya image (Figure \ref{fig12}d; upper frames) is located within the locus of the accretion disk and it represents the asymmetric portion of the disk that was not included in the synthetic spectrum based on a symmetric disk; hence the accretion disk in TT Hya is asymmetric.  An asymmetric disk and gas moving at sub-Keplerian velocities were also discovered in AU Mon from the tomograms (see Figure \ref{fig12}d; lower frames).  A Localized Region was also detected in the AU Mon image; this emission was modeled as a ``spot" and removed from the image (see Figure \ref{fig12}e; lower frames); the density of this region was $8\times10^{-14}$ g\,cm$^{-3}$,  which is comparable to that of the gas stream.
   
In summary, the  {\sc shellspec} code has proven to be an effective tool in the modeling of various accretion structures and for the calculation of critical parameters, such as the mass transfer rate, needed to compute accurate stellar evolution models.
Moreover, the use of the synthetic spectra with tomography has led to the isolation and discovery of asymmetric accretion disks in two disk-state systems which have stable accretion disks.  Moreover, this analysis has led to the successful modeling and extraction of the Localized Region from a composite spectrum, in the case of AU  Mon.

\section{Results}

In this paper, we have derived the behavior of the accretion flows in a range of direct-impact and disk-state Algol binaries with orbital periods from 1.2 to 11.1 days.  These systems undergo mass transfer from a magnetically-active star onto a normal, non-magnetic main sequence star.  We have applied the technique of tomography and a variety of other procedures, both observational and theoretical, to derive a better understanding of the gravitational and magnetic phenomena associated with the mass transfer process in interacting binary stars.   

(1) The $r-q$ diagram was used as a predictive tool to determine which systems will undergo direct-impact accretion and which will form an accretion disk (Figure \ref{fig1}).  In order to understand the predicted dynamical behavior of the gravitational flow along the gas stream, several key dynamical parameters were calculated, namely the thermal, dynamical, and circularization timescales which were then compared with the synchronous velocity, escape velocity for the mass gainer, and the velocity of the gas stream at the impact site.   If only gravitational and Coriolis forces are considered, the gas flow from the L1 point will make direct impact with the surface of the mass gainer in the short-period systems, and the impact can create an equatorial bulge on the surface of this star because the velocity of the flow is typically much higher than the synchronous velocity of the star.  The gas at the impact site will be turbulent, and the tangential component of the flow will propel the gas around the star well beyond the impact site on timescales of hours to days.  The predicted behavior of the gas at the shock site was calculated for both the Cartesian and velocity frames for comparison with the Doppler tomograms (Figure \ref{fig2}).

(2) The image reconstruction technique of back-projection Doppler tomography was described mathematically, along with its advantages and constraints.  In essence, this technique creates velocity images by summing the normalized intensities of the spectra at all velocities across the spectrum and for all viewing positions around the orbit.  This procedure requires the availability of numerous spectra with high spectral resolution and high resolution in orbital phase (Figure \ref{fig3}).  The 2D images shown in Figure 4 were computed without any filtering, smoothing, denoising or other regularization procedures, except for interpolation in orbital phase.  

(3) For the application of tomography, we analyzed over 1300 time-resolved spectra of 13 Algol binaries collected at multiple wavelengths and multiple epochs (Figure 4).  Most of the tomograms were constructed from H$\alpha$ spectra, and we included images derived from the H$\beta$, He I $\lambda$6678, and Si II $\lambda$6371 lines in the optical region, as well as the Si IV $\lambda$1394 line in the ultraviolet region.  The velocity images produced via tomography display seven distinct sources of emission that provide evidence of both gravitational and magnetic effects (Figure \ref{fig5}): 
(i) a source of magnetic activity associated with the cool mass-losing star, including emission from the chromosphere, prominences, flares, and coronal mass ejections; (ii) a gas stream along the predicted gravitational trajectory, which was identified in the images of nearly all of the systems in our survey;  (iii) a star-stream impact region where the gas stream strikes the surface of the mass-gaining star; (iv) a bulge of extended absorption or emission around the mass-gaining star produced by the impact of the high velocity stream onto the slowly rotating photosphere; (v) a transient or permanent accretion disk around the mass-gaining star; (vi) an absorption zone which may be the locus of hotter gas beyond the $10^4$ K regime associated with H$\alpha$ emission; and (vii) a disk-stream impact region where the circling gas strikes the incoming stream and is slowed by the impact to sub-Keplerian velocities.  In addition, (viii) a hot spot where the gas stream strikes an already well-formed disk was found in one system by means of S-wave analysis.

The images, trailed spectrograms, and representative spectra at quadrature provide us with a consistent progression from the stream state to the disk state when the binaries are displayed in order of increasing orbital separation (Figure \ref{fig6}).  This progression reveals the existence of the Alternating Systems which have been found in both the stream state and the disk state at different epochs.  In addition, the multiwavelength spectra provide us with views of the gas flows at different temperatures (e.g., H$\alpha$ vs. He I spectra; Figure \ref{fig7}).

(4)  We examined the influence of various system properties on the resulting 2D velocity images (see Figure \ref{fig8}).  Orbital inclination causes the image size to decrease as the inclination increases from $50^\circ$ to $90^\circ$. The adoption of an inaccurate systemic velocity results in the distortion of the image. Insufficient orbital phase coverage creates radial spoke-like streaks across the diagram which can be rectified if the data are interpolated in orbital phase. Inadequate velocity resolution across the spectrum can also be addressed by interpolation, but the observed spectra used in the analysis already have high resolution and do not require any interpolation. Finally, shadowing, or the presence of partially opaque gas, will influence the passage of light through a medium and this effect can be noticed if we view the binary from opposite directions.  This latter effect was not detected in U CrB or TT Hya (see Figure \ref{fig8}).  However, non-transparent gas was detected through a comparison between the observed and difference spectra of CX Dra.

(5) Finally, we summarized four methods that have been used to extract the physical properties and other characteristics of the gas flows directly from the velocity images: (a) S-wave analysis; (b) creation of simulated velocity tomograms from hydrodynamic simulations; c) sequential extraction of the separate emission sources using synthetic spectra of the disk, gas stream, and other sources of emission such as the Localized Region; and (d) using synthetic spectra with tomography to confirm that the separate accretion structures have been modeled accurately.     

(a) A comparison of the trailed spectrograms with the velocity curves of the stars showed that there are compact structures within the gas flow that nearly follow the velocity curve of one of the stars; the orbital variation of these S-waves can be used to derive the location of the source in both the velocity and Cartesian frames.  Hence, the S-wave analysis provides a means by which the information in the velocity map can be translated onto the Cartesian frame without knowledge of the composite velocity fields in the tomogram (Figures \ref{fig9}, \ref{fig10}). 

(b) The construction of simulated tomograms from hydrodynamic calculations under controlled conditions provide a tool for comparison with the Doppler tomograms derived from the observations, and provide a direct connection between the geometry and dynamics of the accretion flows in both velocity and Cartesian dimensions under these controlled conditions (Figure \ref{fig11}).  The simulated tomograms permitted an interpretation of the alternating behavior of U CrB and U Sge as due to periodic changes in the density of the gas stream, and hence in the mass transfer rate.  Results from a radio flare survey suggest that the timescale of variability in radio flares from magnetically-active Algol binaries is comparable to that of the alternating systems, and hence the alternating behavior may be triggered by magnetic activity on the cool mass-losing star.

(c) The {\sc{shellspec}} code has been used effectively to compute synthetic spectra in moving transparent or non-transparent 3D structures such as a disk, stream, spot, jet, or shell, in addition to the stars.  It can be used to sequentially model the composite synthetic spectra of combinations of these structures (e.g.,  stars, disk, and gas stream) for comparison with the observed spectra.  
As a result, critical physical parameters, including the mass transfer rate, of these structures have been obtained for TT Hya and AU Mon with this code. These parameters are needed to compute accurate stellar evolution models.   

(d) In combination with tomography, the synthetic spectra have been used to visually confirm the quality of the computed model. This procedure has led to the discovery of asymmetric accretion disks in TT Hya and AU Mon, as well as the isolation of a sub-Keplerian velocity region and a Localized Region in AU Mon.

\section{Conclusions}

The multi-tiered analysis performed in this work involved both theoretical tools and spectroscopic observations of Algol binaries from the stream state to the disk state.  This analysis has provided observational evidence of the effects of gravitational forces on the accretion flows in these interacting binaries.  Moreover, we have used images derived from 2D Doppler tomography to identify associations with magnetic activity of the cool star, which is the source of the accretion flow.  These associations are distinct from the effects of gravitational effects on the gas flows and demonstrate that magnetic effects cannot be ignored in any of  these interacting binary star systems. 

The evidence is in the form of 
(i) chromospheric emission with the same velocity as that of the mass-losing star arising from prominences, flares, and coronal mass ejections, and this emission was detected in 11 of the 13 systems studied;   (ii) detection of a gas stream deflected from the predicted gravitational trajectory (e.g., TT Hya);  and (iii) alternating behavior between stream state and disk state possibly caused by flaring activity (e.g., U Sge, U CrB).  

The extension to 3D Doppler tomography is beginning to provide additional evidence of the influence of magnetic activity on the gas flows beyond the central orbital plane of the binary.  The spectra employed in this study were used to  calculate 3D tomograms of  three systems: U CrB, RS Vul, and $\beta$ Per \citep{agafonovetal06,agafonovetal09,richardsetal10,richardsetal12}.
These 3D images suggest that magnetic threading of the gas, the superhump phenomenon, may alter the direction and intensity of the gas flow in the cases of RS Vul and $\beta$ Per, and may be responsible for deflecting the gas stream away from the central plane to produce a tilted accretion disk in U CrB \citep{richardsetal12}.  These magnetic effects were mimicked using 3D hydrodynamic simulations by boosting the magnitude of the gas stream velocity to Mach 30 and by deflecting the initial gas stream at the L1 point by 45$^\circ$ from the orbital plane in U CrB and RS Vul  \citep{raymer12}. These simulations concluded that a deflected stream is a viable mechanism for producing the strong out-of-plane flows seen in the 3D tomographic images.   

So, we have more to discover from the application of tomography to the rich set of spectra described in this study.

\acknowledgements
We thank the referee for helpful comments on the manuscript. This research involved undergraduate students (JF, MC) and a graduate student (AC), and it was partially supported by National Science Foundation grants AST-0908440 and DMS-1309808, and by the Institut f\"ur Angewandte Mathematik, Ruprecht-Karls-Universit\"at Heidelberg, Germany.


\begin{thebibliography}

\bibitem[Agafonov et al.(2006)]{agafonovetal06}
Agafonov, M. I., Richards, M. T., \& Sharova, O. I. 2006, \apj, 652, 1547	

\bibitem[Agafonov et al.(2009)]{agafonovetal09} 
Agafonov, M.~I., Sharova, O.~I., \& Richards, M.~T.\ 2009, \apj, 690, 1730

\bibitem[Atwood-Stone et al.(2012)]{atwood-stoneetal12}
Atwood-Stone, C., Miller, B. P., Richards, M. T., Budaj, J., \& Peters, G. J. 2012,  \apj, 760, 134 

\bibitem[Budaj \& Richards(2004)]{budaj+richards04}
Budaj, J., \& Richards, M. T., 2004, Contrib. Astron. Obs. Skalnat{\'e} Pleso, 34, 167	

\bibitem[Budaj, Richards \& Miller(2005)]{budajetal05} 
Budaj, J., Richards, M.T., \& Miller, B. 2005, \apj, 623, 411 	

\bibitem[Cugier \& Molaro(1984)]{cugier+molaro84}
Cugier, H., and Molaro, P. 1984, \aap, 140, 105.

\bibitem[Etzel \& Olson(1993)]{etzel+olson93}
Etzel, P. B. \& Olson, E. C. 1993, \aj, 106, 1200

\bibitem[Girard et al.(2011)]{girardetal11}
Girard, M. J. A., Strouthidis, N. G., Ethier, C. R., \& Mari, J. M. 2011, Invest. Ophthalmol. Vis. Sci., 52 (No. 10), 7738

\bibitem[Glazunova et al.(2011)]{glazunovaetal11}
Glazunova, L. V., Mkrtichian, D. E., \& Rostopchin, S. I. 2011, \mnras, 415, 2238
	
\bibitem[Guinan \& Gim{\'e}nez(1994)]{guinan+gimenez94}
Guinan, E. F., \& Gim{\'e}nez, A. 1994, in The Realm of
Interacting Binary Stars, ed. J. Sahade, G. E. McCluskey, Jr. \& Y.
Kondo, (Dordrecht: Kluwer), p. 51

\bibitem[Iben \& Tutukov(1985)]{iben+tutukov85}
Iben, I., \& Tutukov, A. V. 1985, \apjs, 58, 661

\bibitem[Kaitchuck et al.(1994)]{kaitchucketal94}
Kaitchuck, R. H., Schlegel, E. M., Honeycutt, R. K., Horne, K., Marsh,
T. R., White, J. C., \& Mansperger, C. S. 1994, \apjs, 93, 519

\bibitem[Kempner \& Richards(1999)]{kempner+richards99}
Kempner, J. C. \& Richards, M. T. 1999, \apj, 512, 345

\bibitem[Komzik et al.(2008)]{komziketal08}	
Komzik, R., Chochol, D. \& Grygar, J. 2008, Contrib. Astron. Obs. Skalnat{\'e} Pleso, 38, 538

\bibitem[Lubow \& Shu(1975)]{lubow+shu75}
Lubow, S. H., \& Shu, F. H. 1975, \apj, 198, 383

\bibitem[Marsh(2001)]{marsh01}
Marsh, T. R. 2001, \textit{Lecture Notes in Physics}  573, 1


\bibitem[Marsh \& Horne(1988)]{marsh+horne88}
Marsh, T. R. \& Horne, K. 1988, \mnras, 235, 269

\bibitem[Miller et al.(2007)]{milleretal07} 
Miller, B. P., Budaj, J., Richards, M. T., Koubsk\'{y}, P. \& Peters, G. J. 2007, \apj, 656, 1075 

\bibitem[Morales-Rueda(2004)]{morales-rueda04}
Morales-Rueda, L. 2004, AN, 325, 193

\bibitem[Olson(1982)]{olson82}
Olson, E. C. 1982, \pasp, 94, 700

\bibitem[Olson(1987)]{olson87}
Olson, E. C. 1987, \aj, 94,1043

\bibitem[{Patterson}(1998)]{patterson98}
Patterson, J. 1998, \pasp, 110, 1132 

\bibitem[{Patterson}(1999)]{patterson99}
Patterson, J. 1999, in {Disk Instabilities in Close Binary Systems}, ed. S. Mineshige
\& J. C. Wheeler ( Tokyo: Universal Academy Press), p. 61

\bibitem[Peters \& Polidan(1984)]{peters+polidan84} 
Peters, G.~J., \& Polidan, R.~S. 1984, \apj, 283, 745

\bibitem[Peterson et al. (2010)]{petersonetal10}
Peterson, W. M., Mutel, R. L., Gudel, M. \& Goss, W. M. 2010, \nat, 463, 207

\bibitem[Peterson et al.(2011)]{petersonetal11} 
Peterson, W.~M., Mutel, R.~L., Lestrade, J.-F., G{\"u}del, M., \& Goss, W.~M. 2011, \apj, 737, 104 

\bibitem[Prinja et al.(2011)]{prinjaetal11}
Prinja, R. K., Long, K. S. Long, Richards, M. T., Witherick, D.K., \& Peck, L.W. 2012, \mnras, 419, 3537 

\bibitem[Radon(1917)]{radon1917}
Radon, J. 1917, {Berichte S{\" a}chsische Akademie der Wissenschaften Leipzig Math. Phys. Kl.} 
69, 262  (reprinted in 1983: {Proceedings of Symposia in Applied Math} 27, 71)

\bibitem[Raymer(2012)]{raymer12} 
Raymer, E.  2012, \mnras, 427, 1702 

\bibitem[Retter et al.(2005)]{retteretal05}
Retter, A., Richards, M. T., \& Wu, K. 2005, \apj, 621, 417

\bibitem[Richards(1992)]{richards92} 
Richards, M.~T. 1992, \apj, 387, 329

\bibitem[Richards(1993)]{richards93}
Richards, M. T. 1993, \apjs, 86, 255

\bibitem[Richards(2004)]{richards04} 
Richards, M.~T. 2004, AN, 325, 229 	

\bibitem[Richards et al.(2012)]{richardsetal12} 
Richards, M.~T., Agafonov, M.~I., \& Sharova, O.~I.\ 2012, \apj, 760, 8 

\bibitem[Richards \& Albright(1993)]{richards+albright93} 
Richards, M. T., \& Albright, G. E. 1993, \apjs, 88, 199

\bibitem[Richards \& Albright(1996)]{richards+albright96} 
Richards, M. T., \& Albright, G. E. 1996, in Stellar Surface Structure, eds. K. Strassmeier and J. Linsky (Dordrecht: Kluwer), 493

\bibitem[Richards \& Albright(1999)]{richards+albright99} 
Richards, M.~T., \& Albright, G.~E. 1999, \apjs, 123, 537

\bibitem[Richards et al.(1995)]{richards+albright+bowles95} 
Richards, M. T., Albright, G. E., \& Bowles, L. M. 1995, \apjl, 438, L103

\bibitem[Richards, Jones \& Swain(1996)]{richards+jones+swain96} 
Richards, M. T., Jones, R. D., \& Swain, M. A. 1996, \apj, 459, 249

\bibitem[Richards et al.(2000)]{richardsetal00} 
Richards, M.~T., Koubsk{\'y}, P., {\v S}imon, V., Peters, G. J., Hirata, R.,  {\v S}koda, P., \& Masuda, S. 2000, \apj, 531, 1003 

\bibitem[Richards \& Ratliff(1998)]{richards+ratliff98}
Richards, M. T., \& Ratliff, M. A. 1998, \apj, 493, 326

\bibitem[Richards et al.(2010)]{richardsetal10} 
Richards, M. T., Sharova, O., \& Agafonov, M. 2010, \apj, 720, 996	

\bibitem[Richards et al.(2003)]{richards+waltmanetal03}
Richards, M. T., Waltman, E. B., Ghigo, F., \& Richards, D. St. P. 2003, \apjs, 147, 337

\bibitem[Sarna et al.(1997)]{sarnaetal97}
Sarna, M. J., Muslimov, A. G., \& Yerli, S. K. 1997, \mnras, 286, 209

\bibitem[Sarna et al.(1998)]{sarnaetal98}
Sarna, M. J.,Yerli, S. K., \& Muslimov, A. G. 1998, \mnras, 297, 760

\bibitem[Schwope et al.(2004)]{schwopeetal04}
Schwope, A. D., Staude, A., Vogel, J., \& Schwarz, R.  2004, AN, 325, 197

\bibitem[Sharova et al.(2012)]{sharovaetal12} 
Sharova, O. I. et al. 2012,  in IAU Symposium 282, From Interacting Binaries to Exoplanets: Essential Modeling Tools, eds. M. T. Richards \& I. Hubeny (Cambridge: Cambridge U. Press),  p. 201

\bibitem[Shepp(1983)]{shepp83}
Shepp, L. A. 1983, Computed Tomography, Proc. of Symposia in  Applied Math., Vol. 27, ed. L. A. Shepp (Providence: American Mathematical Society)

\bibitem[Steeghs(2004)]{steeghs04}
Steeghs, D. 2004, AN, 325, 185

\bibitem[Struve(1949)]{struve49}
Struve,O. 1949, \mnras, 109, 487

\bibitem[Szkody et al.(2000)]{szkodyetal00}
Szkody, P., Desai, V., \& Hoard, D. W. 2000, \aj, 119, 365

\bibitem[Szkody et al.(2001)]{szkodyetal01}
Szkody, P., Nishikida, K., Long, K. S. \& Fried, R. 2001, \aj, 121, 2761

\bibitem[Vesper \& Honeycutt(2001)]{vesper+honeycutt93}
Vesper, D. N., \& Honeycutt, R. K. 1993, \pasp, 105, 731

\bibitem[Vesper et al.(2001)]{vesperetal01}
Vesper, D., Honeycutt, K., \& Hunt, T. 2001, \aj, 121, 2723

\bibitem[Vrtilek et al.(2004)]{vrtileketal04}
Vrtilek, S. D., Quaintrell, H., Boroson, B., \& Shields, M. 2004, AN, 325, 209

\bibitem[Vogt(1982)]{vogt82}
Vogt, N. 1982, \apj, 252, 653

\bibitem[Wheeler(2007)]{wheeler07}
Wheeler, J. C. 2007, {Cosmic Catastrophes} (Cambridge: Cambridge Univ. Press)

\bibitem[Zavala et al.(2010)]{zavalaetal10}
Zavala, R. T., Hummel, C. A., Boboltz, D. A., Ojha, R. Shaffer, D. B., Tycner, C., Richards, M. T.  \& Hutter, D. J.  2010, \apjl, 715, L44

\end{thebibliography}
\end{document}